\documentclass[longauth]{aa} 
\usepackage[colorlinks=true,linkcolor=blue,citecolor=blue]{hyperref}

\usepackage{graphicx}
\bibpunct{(}{)}{;}{a}{}{,} 
\usepackage{multirow}
\usepackage{lipsum,multicol}
\usepackage{natbib}
\usepackage[varg]{txfonts}
\usepackage{orcidlink}

\def \kms{km\,s$^{-1}$}
\def \ms{m\,s$^{-1}$}

\def \mearth{$M_\oplus$}
\def \rearth{$R_\oplus$}
\def \mjup{$M_{\rm Jup}$}
\def \rjup{$R_{\rm Jup}$}
\def \Hei{\ion{He}{I}}
\def \Sii{\ion{Si}{I}}
\def \Ha{H$\alpha$}
\def \nuno{Night~1}
\def \ndue{Night~2}
\def \ntre{Night~3}
\def \nquattro{Night~4}
\def \sun{\odot}

\def\parvati{{\fontfamily{pcr}\selectfont PARVATI}}
\def\logrhk{\ensuremath{\log R^{\prime}_{\mathrm{HK}}}}
\def\planet{V1298\,Tau\,b}
\def\starr{V1298\,Tau}

\begin{document}

        \title{The GAPS Programme at TNG}
        \subtitle{LXXXI. The puzzle of the \Hei\ absorption signal in the atmosphere of \planet\ with GIARPS@TNG}
        
        \titlerunning{\ion{He}{I} in \planet\ with GIANO-B@TNG}
        \authorrunning{G. Guilluy et al.}
        
        \author{G. Guilluy \orcidlink{0000-0002-1259-2678}\thanks{Corresponding author: \email{gloria.guilluy@inaf.it}}\inst{1}
                \and
                J.~Orell-Miquel \orcidlink{0000-0003-2066-8959}\inst{2}
                \and
                A.~Sozzetti\orcidlink{0000-0002-7504-365X}\inst{1} 
                \and
                M.~C~D'Arpa\inst{3} 
                \and
                N.~Greco\inst{4,5,6} 
                \and
                A.~F.~Lanza\inst{4} 
                \and
                D.~Sicilia\inst{4} 
                \and
                A.~Caldiroli\inst{7} 
                \and
                R.~Spinelli\inst{3,7} 
                \and
                D.~Locci\inst{3} 
                \and
                P.~Leonardi\inst{8,9} 
                \and
                L.~Mancini\orcidlink{0000-0002-9428-8732}\inst{10,1} 
                \and
                S.~Colombo\inst{3} 
                \and
                M.~Rainer\inst{11} 
                \and
                F.~Borsa\inst{11} 
                \and
                A.~Maggio\inst{3} 
                \and
                A.~S.~Bonomo\orcidlink{0000-0002-6177-198X}\inst{1} 
                \and
                L.~Affer\inst{3}
                \and
                V.~Andretta\inst{12}
                \and
                F.~Biassoni\inst{11} 
                \and
                A.~Bignamini\inst{13} 
                \and
                W.~Boschin\inst{14}
                \and
                M~Brogi\inst{15}
                \and
                C.~Di~Maio\inst{3}
                \and
                L.~Fossati\inst{16}
                \and
                E.~Gallo\inst{17} 
                \and
                F.~Haardt\inst{18} 
                \and
                M.~López-Puertas\inst{19}
                \and
                V.~Lorenzi\inst{14}
                \and
                L.~Malavolta\inst{8,9}
                \and
                J.~Maldonaldo\inst{3}
                \and
                F.~Manni\orcidlink{0009-0003-1841-7381}\inst{1,10}
                \and
                G.~Micela\inst{3} 
                \and 
                F.~Murgas\orcidlink{0000-0001-9087-1245}\inst{20,21}
                \and
                V.~Nascimbeni\inst{9} 
                \and
                E.~Palle\inst{20,21}
                \and
                J.~Sanz-Forcada\inst{22}
                \and
                G.~Scandariato\inst{4}
                \and 
                J.~Southworth\inst{23}
                \and
                J.~Tregloan-Reed\inst{24}
        }
        
        \institute{
                INAF -- Osservatorio Astrofisico di Torino, Via Osservatorio 20, 10025, Pino Torinese, Italy
                \and
                Department of Astronomy, University of Texas at Austin, 2515 Speedway, Austin, TX 78712, USA 
                \and 
                INAF -- Osservatorio Astronomico di Palermo, P.zza Parlamento 1, 90134 Palermo, Italy 
                \and 
                INAF -- Osservatorio Astrofisico di Catania, Via S. Sofia 78, 95123, Catania, Italy 
                \and
                Dipartimento di Fisica e Astronomia ``Ettore Majorana'', Università di Catania, Via S. Sofia 64, 95123 Catania, Italy 
                \and
                Scuola Superiore di Catania, Università di Catania, Via Valdisavoia 9, 95123 Catania, Italy 
                \and
                Como Lake Center for Astrophysics (CLAP), DiSAT, Università degli Studi dell’Insubria, via Valleggio 11, 22100 Como, Italy 
                \and 
                Dipartimento di Fisica e Astronomia, Università degli Studi di Padova, Vicolo dell’Osservatorio 3, 35122 Padova, Italy 2  
                \and 
                INAF-- Osservatorio Astronomico di Padova, Vicolo dell’Osservatorio 5, 35122 Padova, Italy 
                \and
                Department of Physics, University of Rome ``Tor Vergata'', Via della Ricerca Scientifica 1, 00133, Roma, Italy 
                \and
                INAF -- Osservatorio Astronomico di Brera, Via E. Bianchi 46, 23807, Merate (LC), Italy
                \and 
                INAF -- Osservatorio Astronomico di Capodimonte, Salita Moiariello 16, 80131, Naples, Italy 
                \and
                INAF -- Osservatorio Astronomico di Trieste, via Tiepolo 11, 34143, Trieste, Italy 
                \and
                INAF -- Fundación Galileo Galilei, Rambla José Ana Fernandez Pérez 7, 38712 Breña Baja, Tenerife, Spain 
                \and
                Department of Physics, University of Turin, Via Pietro Giuria 1, 10125 Torino, Italy 
                \and
                Space Research Institute, Austrian Academy of Sciences, Schmiedlstrasse 6, 8042, Graz, Austria 
                \and
                Department of Astronomy,University of Michigan,1085 S University, Ann Arbor, MI 48109, USA 
                \and
                Center for Astrophysics and Space Science, NYUAD Research Institute, New York University Abu Dhabi, PO Box 129188, Abu Dhabi, UAE 
                \and
                Instituto de Astrofísica de Andalucía (IAA-CSIC), Glorieta de la Astronomía s/n, 18008 Granada, Spain 
                \and
                Instituto de Astrof\'{i}sica de Canarias (IAC), 38205 La Laguna, Tenerife, Spain 
                \and
                Departamento de Astrof\'isica, Universidad de La Laguna (ULL), 38206 La Laguna, Tenerife, Spain 
                \and
                Centro de Astrobiología, CSIC-INTA, Camino bajo del Castillo s/n,
                28692 Villanueva de la Cañada, Madrid, Spain 
                \and
                Astrophysics Group, Keele University, Staffordshire, ST5 5BG, United Kingdom 
                \and
                Centro de Astronomía de la Universidad de Antofagasta, Angamos 601, Antofagasta, 1270300, Chile 
        }
        
        \date{Received 15 July 2026 / Accepted  23 October 2026
        }
        \abstract 
        {
                Atmospheric escape shapes the mass, radius, and atmospheric composition of close-in planets, primarily over their first tens of Myr. The young \starr\ system (10\,--\,30~Myr) represents a benchmark for investigating photoevaporation and planetary evolution at early stages of a system's existence.}
        {The metastable \Hei\ near-infrared triplet is a key proxy for exoplanetary atmospheric escape. We aim to detect and characterise \Hei\ in the extended atmosphere of \planet.}
        {We analysed four transits spanning a four-year baseline using the GIARPS (GIANO-B + HARPS-N) observing mode at the Telescopio Nazionale \textit{Galileo}. We re-derived the projected spin-orbit angle by modelling the Rossiter-McLaughlin effect. Using the simultaneous optical and near-infrared high-resolution transmission spectroscopy of \planet, we monitored the \Hei\ triplet alongside key chromospheric indicators (\logrhk, \Ha, and \ion{Na}{i}) to disentangle stellar activity from planetary evaporation.}
        {While the stellar spectra show significant night-to-night variability, transmission spectroscopy reveals persistent \ion{He}{i} absorption across all four epochs. This contrasts with optical indicators (\Ha\ and \ion{Na}{i}), observed over the two nights with simultaneous optical monitoring, which fluctuate between emission and absorption due to stellar activity. The consistent presence of \ion{He}{i} over these four years could suggest a planetary origin, despite stellar-driven depth modulation. \textsc{ATES} hydrodynamic simulations show that stellar XUV flux variations significantly modulate the \Hei\ absorption depth and equivalent width. Furthermore, stellar wind fluctuations likely drive the observed night-to-night variability of the exospheric structure.}
        {The detection of \Hei\ across all epochs could support the assumption of planetary evaporation, despite it being modulated by stellar activity. Our results underscore the fact that disentangling a planetary signal from stellar contamination remains a big challenge for highly active hosts. Ultimately, confirming the planetary nature of the \Hei\ signature requires comprehensive activity monitoring and full transit coverage to establish a robust baseline.}
        \maketitle

        \section{Introduction}
        Understanding exoplanet evolution is crucial to explaining system architectures and atmospheric stability. A key feature is the so-called radius gap (or evaporation valley) around $\sim$1.8~\rearth, separating dense, rocky super-Earths from larger, volatile-rich mini-Neptunes \citep[e.g.][]{Fulton2017, Fulton2018}. Its origin likely involves both formation and evolutionary processes: migration could shape the sub-Neptune peak, while photoevaporation can produce the super-Earth population \citep{Burn2024}. Alternatively, mechanisms such as core-powered mass loss could produce the observed bimodal distribution on their own (e.g. \citealt{Gupta2019,Loyd2020}). Studying atmospheric escape in planets near the radius gap makes it possible to test whether evaporation drives these demographics and thereby understand the transition from gas-rich planets to stripped cores, especially in young systems where escape is most efficient (\citealt{sanz-forcada2011} and references therein).
        \\
        \indent{The} primary driver of atmospheric erosion is stellar X-ray and extreme ultraviolet (XUV) irradiation \citep[e.g.][]{Lammer2003, lecavelier2004}. As a star ages and spins down, its magnetic activity and XUV output decrease significantly \citep[e.g.][]{Owen2019}. Consequently, for most exoplanets orbiting mature stars, the atmospheric evaporation responsible for the radius gap has largely concluded. Planets around young stars ($\lesssim 100$~Myr) therefore offer a unique opportunity to observe atmospheric escape in action during its most transformative epoch.

In this context, the metastable helium triplet at 10832.06\,\AA, 10833.22\,\AA, and 10833.31\,\AA\ (vacuum wavelength \Hei, e.g. \citealt{Seager2000, Oklop2018, Spake2018}) has emerged as a premier diagnostic of escaping atmospheres. The \Hei\ $\lambda$10833\,\AA
        triplet is an important astrophysical diagnostic, tracing neutral helium in its triplet state. It forms mainly through photoionisation by ultraviolet radiation ($\lambda$\,<504~\AA), followed by recombination into the metastable orthohelium 2$^3$S state. Once in this state, He atoms can absorb incoming photospheric near-infrared photons, producing the \Hei\,10833~\AA\ absorption feature. Unlike traditional tracers such as Ly$\alpha$, the \Hei\ triplet is less sensitive to interstellar medium absorption and geocoronal contamination. Furthermore, its accessibility from the ground enables high-resolution instruments to probe the dynamics of planetary outflows in detail \citep[e.g.][]{Nortmann2018, Allart2018}.

        The \starr\ planetary system is a key laboratory for studying atmospheric photoevaporation in the early stages of planetary evolution. The host star is a young ($\sim$10–30~Myr, \citealt{David2019a, Suarez2021}), rapidly rotating ($P_\mathrm{rot}\approx$2.91~days, \citealt{Finocietyetal23}) K-type dwarf. Its strong magnetic activity complicates radial velocity (RV) measurements \citep[e.g.][]{Suarez2021, maio2023,Damasso2026}, but provides an ideal environment for studying extreme irradiation effects on young planetary atmospheres. 
        Among its four known transiting planets \citep{David2019a}, \planet\ ($P$\,$\sim$\,24.14~days, $R \sim 9.9$~\rearth; Table~\ref{tab:master_parameters}) is a prime target for transmission spectroscopy. Hubble Space Telescope \citep{Barat2024} and James Webb Space Telescope \citep{Barat2025} observations revealed an H/He-dominated atmosphere, detecting $\text{CO}_2$, $\text{H}_2\text{O}$, $\text{CO}$, $\text{CH}_4$, $\text{SO}_2$, and $\text{OCS}$, alongside a high internal temperature, $T_\mathrm{int}$, of  $\sim$500~K. Specifically, $T_\mathrm{int}$ is a parameter of the temperature-pressure profile model (from \citealt{ParmentierGuillot2014}, as adopted by \citealt{Barat2025}) that accounts for the intrinsic internal heat flux. These studies imply a low planetary mass of 12–15~\mearth, independently confirmed by \citet{Livingstonetal26}, whose dynamical estimate
        of $\sim$13.1~\mearth\ was derived by modelling the transit timing variations (TTVs). These values are drastically lower than previous RV estimates (e.g. $M_\mathrm{P} = 203 \pm 60$~\mearth, \citealt{Suarez2021}), and indicate
        that \planet\ may be an inflated gas-dwarf progenitor \citep{Barat2025} of mature sub-Neptunes.
        Despite predictions of ongoing atmospheric escape driven by the host star's intense XUV irradiation \citep{Poppenhaeger2021, maggio_new_2022}, \Hei\ observations of \planet\ remain inconclusive. Previous campaigns yielded either non-detections from partial transits \citep{vissapragada_search_2021} or ambiguous absorption signals blended with stellar variability or the transit of planet~d \citep{gaidos_zodiacal_2022}. This discrepancy between predictions and observations could stem from four main challenges: TTV-induced ephemeris uncertainties, the difficulty of capturing the long transit ($\sim$7~hours) from a single ground-based site, V1298\,Tau\,d atmospheric contamination, and the strong, highly variable stellar \Hei\ background originating in the star's active chromosphere \citep[e.g.][]{Guilluy2020}. In young, active stars such as \starr, stellar signals can mimic or mask planetary absorption, creating a puzzle that is difficult to resolve without simultaneous monitoring of activity indicators.

        In this work, we used high-resolution GIANO-B observations of the metastable helium triplet to investigate the atmospheric outflow of \planet\ while carefully accounting for its complex stellar environment. Our observational setup avoided simultaneous planet~d transiting (Sect.~\ref{sec_observations}), preventing spectral contamination, and guaranteeing an unambiguous analysis of \planet's atmospheric signal. The paper is organised as follows: Sect.~\ref{sec_observations} describes the observations, while Sect.~\ref{system_par} includes an analysis of the Rossiter-McLaughlin effect, the data reduction, and our findings, which are presented and contextualised within the framework of stellar activity. In Sect.~\ref{Discussion}, we discuss the role of star-planet interactions in shaping the \Hei\ signature, contextualize our results within previous literature, and describe our modelling of the planetary evolution.  We summarise our conclusions in Sect.~\ref{sec_conclusion}.

        \section{Observations} \label{sec_observations}

        We observed four transits of \planet, namely UT 28 January 2020 (\nuno), UT 07 December 2020 (\ndue), UT 07 October 2024 (\ntre), and UT 31 October 2024 (\nquattro). A log of the observations is provided in Table~\ref{log}.
        Observations were acquired exploiting the GIARPS observing mode \citep{GIARPS_claudi} at the Telescopio Nazionale Galileo (TNG). GIARPS enables simultaneous high-resolution observations in both the visible (0.39–0.69~$\mu$m) with HARPS-N ($R \approx 115{,}000$) and the near-infrared (0.95–2.45~$\mu$m) with GIANO-B ($R \approx 50{,}000$). 
        During \ntre\ and 4, some instrumental problems occurred that prevented us from using the GIARPS mode for the entire observing night.  Since \Hei\ was the main focus of our research, it was decided to prioritise observations with GIANO-B. Therefore, we only have a few HARPS-N pre-transit and ingress spectra for \ntre, and a few in-transit spectra for \nquattro. 
        
        For the GIANO-B observations, we applied an ABAB nodding pattern \citep{GIARPS_claudi} to improve thermal background noise subtraction. GIANO-B is an echelle spectrograph covering four near-infrared bands ($Y$, $J$, $H$, $K$), divided into 50 orders. 
        The HARPS-N observations were performed with the B-fibre positioned on the sky. This instrument, also an echelle spectrograph, provides spectral coverage across 69 distinct orders for each fibre. 
        \\
        \indent{Based on the TTV dynamical model from Leonardi in prep., we calculated the orbital phases of all known planets in the system for each observing run. None of the other planets were transiting during our observations, ensuring that our GIANO-B and HARPS-N spectra are free from atmospheric contamination by secondary planets (Fig.~\ref{orbite}).}
        \nuno\ and \nquattro\ comprise observations taken before and during the early phases of the transit, while \ndue\ and \ntre\  cover spectra acquired during and after the transit. During \nquattro, the target was not fully in focus for some observations. GIANO-B experienced guiding problems around 01:30 UT (orbital phase $\sim$\,$-0.0022$), but the tracking stabilised by 01:52 UT (orbital phase $\sim$\,$0.0029$), allowing observations to continue.\\
        \indent{Furthermore}, we observed that on \ntre\ and 4, towards the end of
        the night, specifically from UT 04:45:06 on \ntre\ and UT 05:12:46 on
        \nquattro\ until the end of the night, an emission line appears in the
        GIANO-B raw spectra at the position of \Hei\ in the Earth's rest frame
        (Fig.~\ref{emission_line}). Unfortunately, on these two
        nights the difference between the systemic velocity ($v_{\rm sys}$) and the
        barycentric Earth radial velocity ($v_{\rm{berv}}$) is small (on average
        $\sim$$4.2$~\kms\ and $\sim$$-5.8$~\kms\ for \ntre\ and 4, respectively), so
        this emission feature falls directly over the stellar helium line. The line
        corresponds to the metastable He($2^3S$) airglow. First detected in 1959
        \citep{Mironov1959}, it probes the Earth's upper thermosphere via
        solar-induced fluorescence of He($2^3S$), peaking at altitudes above 250~km
        with densities of a few atoms cm$^{-3}$ \citep{Waldrop2005}. Crucially,
        because this emission arises just before nautical twilight and grows
        progressively stronger over time, the standard A-B (B-A) nodding subtraction
        cannot remove it effectively. For our analysis, we therefore masked the
        pixels affected by this feature in the individual images. We emphasise that
        for future observations, especially those taken near twilight, the
        difference between $v_{\rm sys}$ and $v_{\rm{berv}}$ should be kept large
        enough to prevent this airglow from contaminating the stellar helium line.

        \begin{figure}
                \centering
                \includegraphics[width=0.95\linewidth]{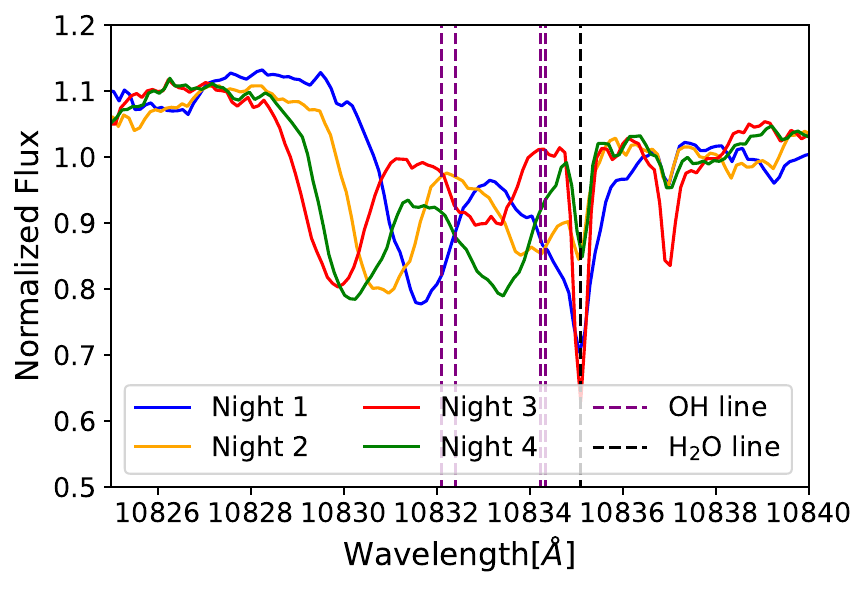}
                \caption{Time-averaged spectra for each night in the telluric rest frame, with the H$_2$O telluric absorption line at approximately 10835.1~\AA\ and the OH telluric emission doublets shown in black and purple, respectively. For clarity, the averaged spectra are displayed after being normalised by their median value.}
                \label{rest_frame}
        \end{figure}

        \begin{table*}
                
                \centering
                \caption{Log of observations.} \label{log}
                \resizebox{17cm}{!}{
                        \begin{tabular}{c c c c  c  c   c c  l l }
                                \hline \hline
                                \multicolumn{2}{c}{Date} & Instrument & $N_{\mathrm{obs}}$ & Exp. time & Airmass & S/N & PI Program & PI Name\\         
                                &   [UT]   &       &                             &           [s]            & min$\rightarrow$mean$\rightarrow$max      &  min$\rightarrow$mean$\rightarrow$max  & &     \\
                                \hline
                                \multirow{2}{*}{\nuno} & \multirow{2}{*}{2020-01-28}  & GIANO-B    & 56[55]                    & 300                   & \multirow{2}{*}{1.01$\rightarrow$1.21$\rightarrow$2.05}   & 26.1$\rightarrow$40.1$\rightarrow$52.0 & GAPS2 & G. Micela\\
                                & & HARPS-N    & 47[47]                    & 400                   &                                   & 23.1$\rightarrow$39.1$\rightarrow$50.9 & \\ 
                                \noalign{\medskip}
                                \multirow{2}{*}{\ndue} & \multirow{2}{*}{2020-12-07}  & GIANO-B    & 80[72]                    & 300                   & \multirow{2}{*}{1.01$\rightarrow$1.21$\rightarrow$2.07}   & 26.0$\rightarrow$44.2$\rightarrow$53.82 & GAPS2 & G. Micela \\ 
                                & & HARPS-N    & 71[71]                    & 400                   &                                   & 11.1$\rightarrow$28.7$\rightarrow$56.5 & \\  
                                \noalign{\medskip}
                                \ntre & 2024-10-07  & GIANO-B    & 58[53]                    & 300                   & 1.01$\rightarrow$1.12$\rightarrow$1.57   & 29.8$\rightarrow$41.4$\rightarrow$48.8 & CAT24B\_81 & J.~Orell~Miquel\\
                                \noalign{\medskip}
                                \nquattro& 2024-10-31  & GIANO-B    & 76[72]                    & 300                  & 1.01$\rightarrow$1.25$\rightarrow$1.83 & 30.7$\rightarrow$39.9$\rightarrow$48.0 & CAT24B\_81 & J.~Orell~Miquel\\
                                \hline 
                        \end{tabular}
                }
                \tablefoot{Columns denote the observing night, instrument, total number of acquired spectra (with the analysed subset in brackets), exposure time, airmass, signal-to-noise ratio (S/N), and Principal Investigator (PI) program and name. S/N values are computed at order \#39 (\Hei) for GIANO-B, and order \#64 (H$\alpha$) for HARPS-N. For GIANO-B, spectra with S/N below 25 were discarded. Airmass and S/N for HARPS-N observations during Nights 3 and 4 are omitted, as the low number of spectra precluded the H$\alpha$ analysis (Sect.~\ref{sec_observations}).}
        \end{table*}

        \section{Analysis and results} \label{system_par}
        \subsection{Rossiter–McLaughlin effect analysis}\label{RML}  
        
\begin{figure}
                \centering
                $\begin{array}{c}
                        \includegraphics[width=0.95\linewidth,height=12cm]{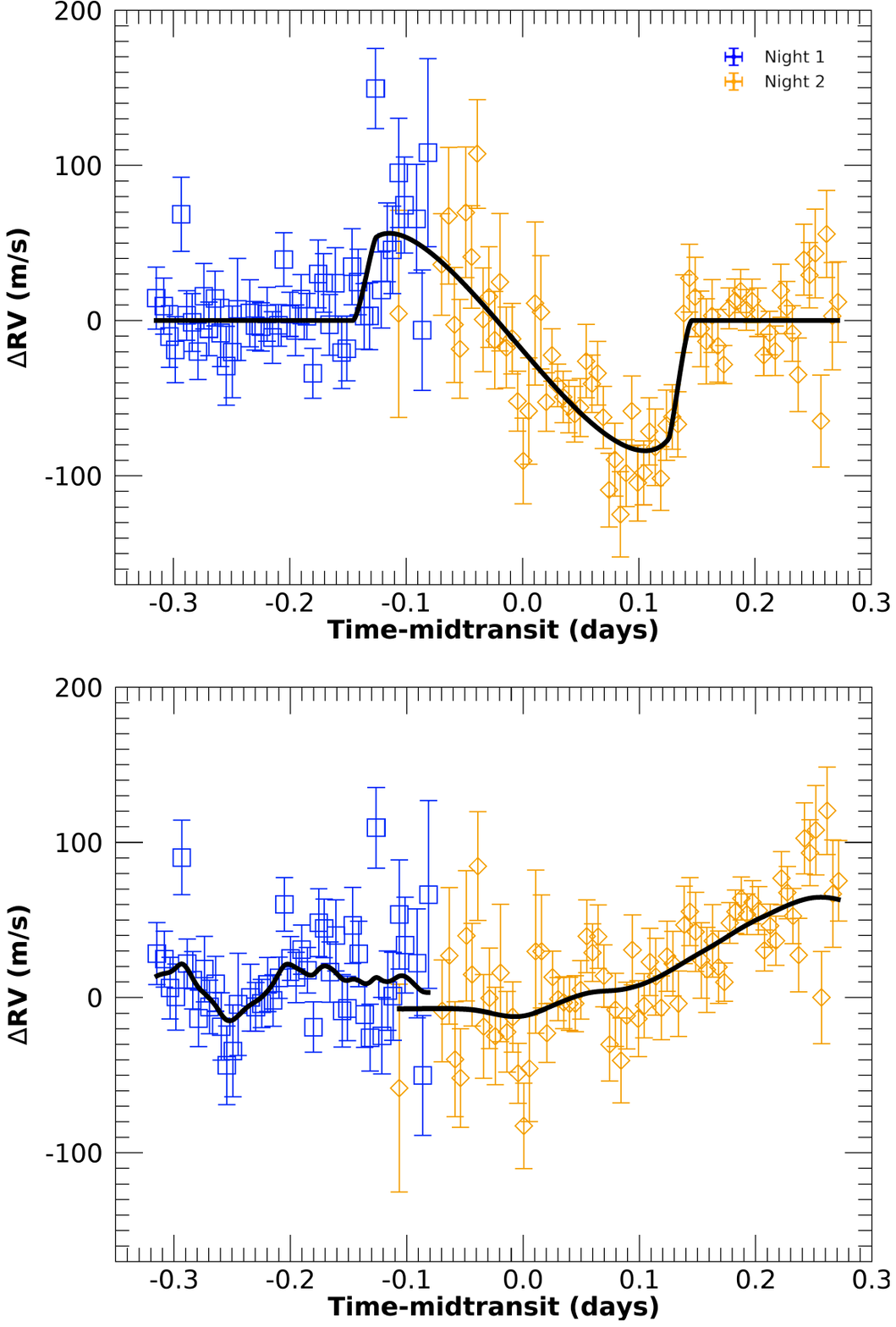}
                \end{array} $
                \caption{\textit{Top}: Phase-folded HARPS-N RV time-series for night 1 (blue squares) and night 2 (yellow diamonds). The contribution from stellar activity has been removed. The thick black line shows the RML component of the best-fit model. \textit{Bottom}: Same but with the contribution from the RML effect removed. The thick black line shows the GP component of the best-fit model.}
                \label{fig:rml_gp_fit}
        \end{figure}

        We used the HARPS-N spectra gathered on night 1 and night 2 to obtain an
        independent constraint on the stellar obliquity of \starr\ relative to the
        normal of planet b's orbital plane, and to measure two new, accurate values
        of the mid-transit time ($T_0$) at these two epochs. The latter is essential
        given that \planet\ is known to exhibit large TTVs due to mutual
        perturbations from the other planets in the system.
        Previous work has already
        detected the RV anomaly produced by the Rossiter-McLaughlin (RML) effect for
        \planet. \citet{gaidos_zodiacal_2022} obtained in-transit spectra during a
        single night with the IRD spectrograph on the 8.2-m Subaru telescope,
        measuring a projected spin-orbit angle of $\lambda = 15^{+15}_{-16}$ deg and
        $\lambda = 2^{+12}_{-4}$ deg from the apparent RV shift and from changes in
        the line profile, respectively. \citet{Johnson2022} observed a partial
        transit with Keck/HIRES (missing egress) and LBT/PEPSI (missing the second
        half and egress), obtaining $\lambda = 4^{+7}_{-10}$ deg from the extracted
        RV sequences for both instruments. The consensus is that \planet\ lies on an
        aligned orbit.\\
        \indent{For} our independent assessment of the spin-orbit alignment of \planet, we
        extracted RVs from the HARPS-N spectra using the template-matching approach
        implemented in the \texttt{TERRA} pipeline \citep{Anglada2012}. For a young, active
        star such as \starr, this approach is preferable to the method based on a
        Gaussian fit to the cross-correlation function (CCF) of the observed spectra
        with a binary mask, as implemented in the standard HARPS-N Data Reduction
        Software pipeline \citep{Dumusque2021}. To model the observed RML signal, we
        adopted the formalism of \citet{Ohta2005}, which is well suited to fitting
        the RML RV signal from template matching. This model was implemented within
        a differential evolution Markov chain Monte Carlo (DEMCMC) algorithm
        \citep{TerBraak2006,Eastman2013}. On both night 1 and night 2, the RV
        sequences show significant variability attributable to stellar activity. We
        included a Gaussian process (GP) framework in the RML modelling to capture
        this correlated RV noise, adopting a Matérn-3/2 covariance kernel
        \citep{Stangret2021}.\\
        \indent{In} the \citet{Ohta2005} formalism, the RML model depends on the following
        parameters: $\lambda$, $T_0$, $P$, the projected stellar rotational velocity,
        $v\sin i_\star$, a linear limb-darkening coefficient, $\epsilon$, the
        inclination angle of the planetary orbit, $i$, the radius ratio,
        $R_\mathrm{p}/R_\star$, and the orbital semi-major axis scaled in units of
        the stellar radius, $a/R_\star$. The common approach is to adopt $\lambda$,
        $T_0$, $v\sin i_\star$, and $\epsilon$ as free parameters, while fixing the
        others to high-precision determinations from the literature (e.g.
        \citealt{Palle2020,Stangret2021}). In the case of \planet, this approach
        risks introducing biases and inconsistencies, primarily because of ($a$) the
        heterogeneity of the methods used to model spectroscopic and photometric
        datasets of variable quality, ($b$) the strong dynamical perturbations in the
        system, which make the use of instantaneous (osculating) orbital elements
        (e.g. $P$) a potentially poor choice, and ($c$) the intrinsic difficulty of
        obtaining consistent estimates of the stellar parameters for the young
        primary. \\
        \indent{Table} \ref{tab:parcomp} compares published stellar and
        planetary parameters drawn from a range of literature sources, including
        \citet{David2019a} (D19a), \citet{David2019b} (D19b),
        \citet{gaidos_zodiacal_2022} (G21), \citet{vissapragada_search_2021} (V21),
        \citet{Feinstein2021} (F21), \citet{feinstein_v1298_2022} (F22),
        \citet{Johnson2022} (J22), \citet{Suarez2021} (SM21),
        \citet{Finocietyetal23} (F23), \citet{sikora2023} (S23), and
        \citet{Livingstonetal26} (L26). As the table shows, the complete set of
        parameters is not always reported across studies, and discrepancies between
        reported values can reach or exceed the 3$\sigma$ level, the most
        significant differences being, as expected, those in the $P$ values. Thus, we chose to treat the full set of RML model parameters as free,
        adopting wide uninformative priors or broad Gaussian priors that comfortably
        bracket the ranges of literature values to optimise the exploration
        of the parameter space (Table \ref{tab:master_parameters}). The global
        RML+GP model contains a total of 14 free parameters, including separate
        values of the epoch of transit centre ($T_{0,1}$, $T_{0,2}$), the GP
        amplitude ($A^{1}_{\mathrm{GP}}$, $A^{2}_{\mathrm{GP}}$), and the GP timescale
        ($\tau^{1}_{\mathrm{GP}}$, $\tau^{2}_{\mathrm{GP}}$) for night 1 and night 2,
        respectively. The priors on $T_{0,1}$ and $T_{0,2}$ were centred on the
        predicted ephemerides (Ephem) derived from the dynamical TTV model of L26,
        while the prior on the linear limb-darkening coefficient was centred on the value obtained with \texttt{LDTk} \citep{Parviainen2015} for a star with the same effective temperature as V1298 Tau and a spectral range corresponding to that of HARPS-N.\\
        \indent{The} global DEMCMC fit converged to a set of posterior distributions whose medians and $1\sigma$ uncertainties (obtained from the $\pm34.13\%$
        intervals of the posteriors) are reported in Table
        \ref{tab:master_parameters}. Figure \ref{fig:rml_gp_fit} shows the
        phase-folded HARPS-N RV time series, with the stellar activity contribution
        removed in the top panel and the RML effect removed in the bottom panel. The
        RML-only and GP-only model components are also overplotted. Overall, the
        central value of $\lambda$ ($31 \pm 14 $~deg) differs from zero at
        the 2.2$\sigma$ level, indicating a possible departure from exact spin-orbit
        alignment. Nevertheless, the large uncertainty on the spin-orbit angle makes
        the retrieved value consistent with previous literature estimates.
        
        \begin{table*}
                \centering
                \caption{Priors and best-fit values of the RML analysis, combined with the stellar and planetary parameters adopted in this work.} \label{tab:master_parameters}
                \footnotesize
                \resizebox{17cm}{!}{
                        \begin{tabular}{l c c l}
                                \hline\hline \\ [-8pt] 
                                Parameter & Prior & Value & Reference \\ [2pt] 
                                \hline \\[-8pt] 
                                \multicolumn{4}{l}{Stellar parameters} \\
                                Effective temperature, $T_\mathrm{eff}$ [K]         & ... & 5050$_{-100}^{+100}$ & SM21 \\ [2pt] 
                                Stellar radius, $R_\star$ [$R_\sun$]                & ... & 1.36$\pm$0.097 & This work$^{(a)}$ \\ [2pt] 
                                Stellar mass, $M_\star$ [$M_\sun$]                  & ... & 1.124$\pm$0.015 & This work$^{(a)}$ \\ [2pt] 
                                Systemic velocity, $v_\mathrm{sys}$ [\kms]          & ... & 16.15$\pm$0.38 & \citet{David2019a} \\ [2pt] 
                                Projected rotational velocity, $v\sin i_\star$ [\kms] & $\mathcal{N}(24.0,2.0)$ & $24.5_{-1.2}^{+1.3}$ & This work \\ [2pt] 
                                Limb darkening coefficient, $\epsilon$              & $\mathcal{N}(\texttt{LDTk},0.2)$ & $0.66_{-0.17}^{+0.16}$ & This work \\ [2pt] 
                                Stellar RV semi-amplitude, $K_{\star}$ [\ms]:       &         &               & \\  [2pt] 
                                & ... & 41$\pm$12 & SM21 \\
                                & ...  & 2.7$\pm$1.1 & This work using ($b$) and $M_\mathrm{p}$/$M_\mathrm{\star}$ from L26\\
                                
                                \vspace{0.01cm} \\
                                \multicolumn{4}{l}{Planetary and Orbital parameters} \\
                                Orbital period, $P$ [days]                          & $\mathcal{N}(\mathrm{Ephem},0.01)$ & $24.1394_{-0.0047}^{+0.0047}$ & This work \\ [2pt] 
                                Transit epoch, $T_{0,1}$ [BJD$_\mathrm{TDB}$]       & $\mathcal{N}(\mathrm{Ephem},0.1)$ & $2458877.639^{+0.012}_{-0.012}$ & This work (\nuno) \\ [2pt] 
                                Transit epoch, $T_{0,2}$ [BJD$_\mathrm{TDB}$]       & $\mathcal{N}(\mathrm{Ephem},0.1)$ & $2459191.4276^{+0.0092}_{-0.0080}$ & This work (\ndue) \\ [2pt] 
                                Transit epoch, $T_{0,3}$ [BJD$_\mathrm{TDB}$]       & ... & 2460591.50759$\pm$0.00059 & Leonardi in prep. (\ntre) \\ [2pt] 
                                Transit epoch, $T_{0,4}$ [BJD$_\mathrm{TDB}$]       & ... & 2460615.6503$\pm$0.0050 & Leonardi in prep. (\nquattro) \\ [2pt] 
                                Scaled semi-major axis, $a/R_\star$                 & $\mathcal{N}(27.0,2.0)$ & $27.35_{-0.94}^{+0.96}$ & This work \\ [2pt] 
                                Radius ratio, $R_\mathrm{p}/R_\star$                         & $\mathcal{N}(0.070,0.005)$ & $0.068_{-0.004}^{+0.004}$ & This work \\ [2pt] 
                                Inclination, $i$ [deg]                              & $\mathcal{N}(89.0,1.0)$ & $89.46_{-0.32}^{+0.31}$ & This work \\ [2pt] 
                                Spin-orbit angle, $\lambda$ [deg]                   & $\mathcal{U}(-180.0,180.0)$ & 31$ \pm$ 14 & This work \\ [2pt] 
                                Planetary mass, $M_\mathrm{p}$ [\mjup]: & & \\      \quad      & ... & 0.63$\pm$0.19 & Derived from $K_\mathrm{\star}$ by SM21 \\
                                & ... & 0.042$\pm$0.017 & Derived from $M_\mathrm{p}$/$M_\mathrm{\star}$ by L26 \\
                                Planet RV semi-amplitude, $K_\mathrm{p}$ [\kms]:    &         &              & \\
                                \quad  & ... & 76.6$\pm$1.2 & This work$^{(c)}$ using $K_\mathrm{\star}$ from SM21 \\
                                \quad  & ... & 76.6$\pm$1.2 & This work$^{(c)}$ using $K_\mathrm{\star}$ derived from L26 \\
                                Eccentricity, $e$                                   & Fixed & 0 & Consistent with a circular orbit (e.g. L26) \\        [2pt] 
                                Transit duration, $T_{14}$ [h]                     & ... & $6.95_{-0.38}^{+0.32}$ & This work$^{(d)}$ \\ [2pt] 
                                Full transit duration, $T_{23}$$^{(4)}$ [h]                & ... & $6.00_{-0.41}^{+0.31}$ & This work$^{(d)}$ \\ [2pt] 
                                Impact parameter, $b$                               & ... & $0.26_{-0.15}^{+0.15}$ & This work$^{(e)}$ \\ [2pt] 
                                Planetary radius, $R_\mathrm{p}$ [\rjup]           & ... & 0.902$\pm$0.060 & This work \\
                                
                                \vspace{0.01cm} \\ 
                                \multicolumn{4}{l}{Noise model and GP parameters} \\
                                RV jitter, $\sigma_{\mathrm{RV}}$ [\ms]             & $\mathcal{U}(0.0,50.0)$ & $23.9_{-5.1}^{+5.0}$ & This work \\ [2pt] 
                                GP amplitude 1, $A^{1}_{\mathrm{GP}}$ [\ms]         & $\mathcal{U}(0.0,50.0)$ & $23.0_{-2.1}^{+2.1}$ & This work \\ [2pt] 
                                GP timescale 1, $\tau^{1}_{\mathrm{GP}}$ [days]     & $\mathcal{U}(0.0,1.0)$ & $0.016_{-0.010}^{+0.011}$ & This work \\ [2pt] 
                                GP amplitude 2, $A^{2}_{\mathrm{GP}}$ [\ms]         & $\mathcal{U}(0.0,50.0)$ & $30.3_{-4.3}^{+4.4}$ & This work \\ [2pt] 
                                GP timescale 2, $\tau^{2}_{\mathrm{GP}}$ [days]     & $\mathcal{U}(0.0,1.0)$ & $0.105_{-0.016}^{+0.020}$ & This work \\ [2pt]
                                \hline
                        \end{tabular}
                }
                \tablefoot{
                        $^{(a)}$ Weighted average of the values reported in the literature.
                        $^{(b)}$ $K_\mathrm{\star} = \left(\frac{2\pi G}{P}\right)^{1/3} \frac{\sin(i)}{\sqrt{1 - e^2}} \frac{M_\mathrm{p}}{(M_\mathrm{p} + M_\star)^{2/3}} $.
                        $^{(c)}$ $K_\mathrm{p} = \frac{K_{\mathrm{\star}} M_{\mathrm{\star}}}{M_\mathrm{p}}$ = $\left(\frac{2\pi G}{P}\right)^{1/3} \frac{\sin(i)}{\sqrt{1 - e^2}} \frac{M_\mathrm{\star}}{(M_\mathrm{p} + M_\star)^{2/3}} $. 
                        $^{(d)}$ T$_\mathrm{14}$= $\frac{P}{\pi} \sin^{-1} \left[ \frac{R_{\star}}{a} \frac{\sqrt{(1+R_{\mathrm{P}}/R_{\mathrm{\star}}))^2 - b^2}}{\sin i} \right]$, $T_\mathrm{23}$=$\frac{P}{\pi} \sin^{-1} \left[ \frac{R_{\star}}{a} \frac{\sqrt{(1-R_{\mathrm{P}}/R_{\mathrm{\star}}))^2 - b^2}}{\sin i} \right]$.
                        $^{(e)}$ $b$=$\frac{a\cos{i}}{R_\star}$.
                }
        \end{table*}

        \subsection{\ntre\ and \nquattro\ $T_{0}$ derivation} \label{sec_midtime}
        Because the HARPS-N spectra obtained during \ntre\ and \nquattro\ were
        insufficient to carry out an RML analysis, we relied on results from the
        ongoing TTV follow-up of the V1298~Tau system (Leonardi in prep.) to
        constrain the time of inferior conjunction. For \ntre, \planet\ was observed
        simultaneously with CHEOPS \citep{Benz_2021} as part of the GO programme
        PR250033 (PI: Leonardi), allowing a precise determination of the transit
        mid-time ($T_{\mathrm{0,3}} = 2460591.50759 \pm 0.00059~\mathrm{BJD_{TDB}}$).
        However, no simultaneous photometric observations were available for night 4,
        preventing a direct measurement of the mid-transit time. In this case, we
        adopted the predicted time of inferior conjunction from the dynamical TTV
        model of Leonardi in prep. ($T_{\mathrm{0,4}} = 2460615.6503 \pm
        0.0050~\mathrm{BJD_{TDB}}$). As a consistency check, we compared the
        mid-transit times derived from the RML analysis for Nights 1 and 2 with the
        predictions of the TTV dynamical model. The two independent determinations
        agree within 1.3$\sigma$, providing additional confidence in the adopted
        ephemeris.

        \subsection{GIANO-B \texorpdfstring{\Hei}{He I} transmission spectroscopy}
        \label{sec_tra}
\begin{figure*}
                \centering
                \includegraphics[width=0.92\linewidth]{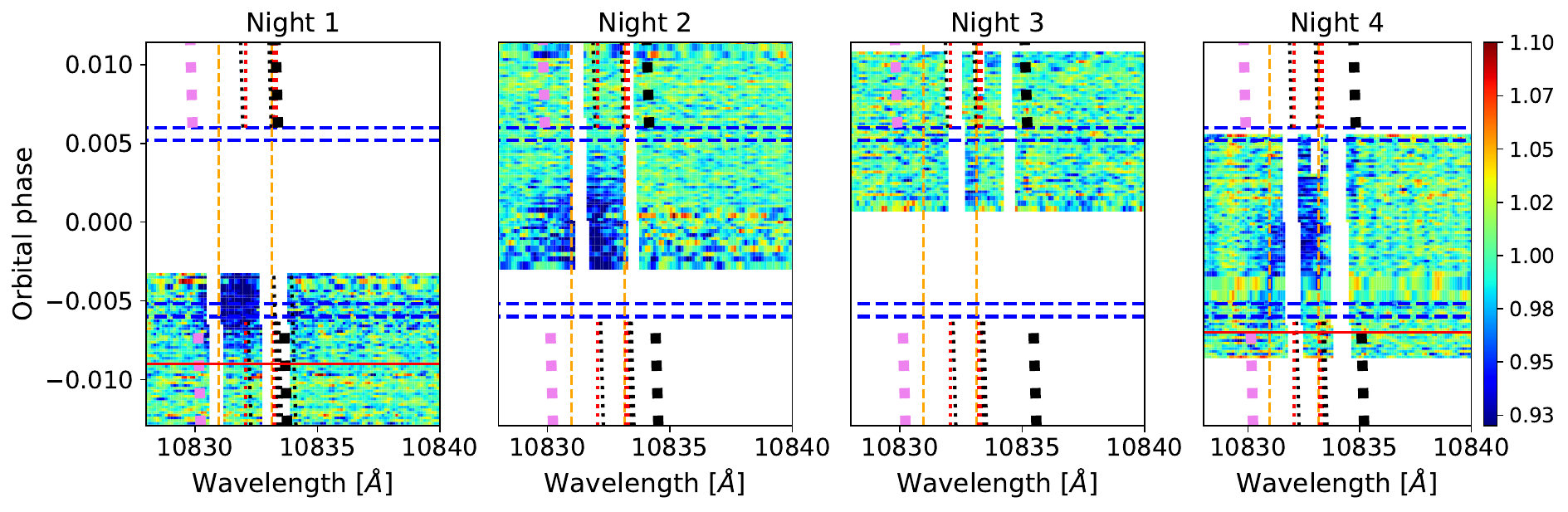}
                \caption{2D maps of transmission spectra in the planet rest frame in the region of the \Hei\ triplet, as a function of wavelength and planetary orbital phase. The contact points $t_1$, $t_2$, $t_3$ and $t_4$ are marked with horizontal dashed blue lines. The regions affected by OH contamination are masked (white vertical bands).  Some residuals are left at the position of the stellar \ion{Si}{I} line ($\sim$10830~\AA, highlighted in pink). This is due to the depth of the line, which can cause difficulties in spectral extraction \citep[e.g.][]{Krishnamurthy2023}. Dotted black and red lines mark the position of the \Hei\ lines in the stellar and planet rest frame, respectively. Lines with black squares mark the position of the strong H$_2$O telluric line at $\sim$10835~\AA. Vertical orange lines indicate the spectral region where the EWs of the individual transmission spectra in Fig.~\ref{HE_ew} have been computed. Orange horizontal line marks the orbital phase used as a threshold to compute the master-out for \nuno\ e \nquattro, respectively.}
                \label{HE_maps}
        \end{figure*}

        To probe the \Hei\ absorption in the extended atmosphere of \planet, we focussed on GIANO-B echelle order \#39. Telluric features, particularly water vapour lines near 10835.1~\AA, were corrected using the ESO {\tt Molecfit} software \citep{Smette2015, Kausch2015}. This spectral region also contains two OH emission doublets close to the \Hei\ triplet: the first at $\sim$10832.1 and $\sim$10832.4~\AA, and the second at $\sim$10834.2 and $\sim$10834.3~\AA\ (vacuum wavelengths). Although the A-B nodding acquisition automatically removes these emission features, variations in atmospheric seeing during the night can leave residual signals. We therefore visually identified the positions of these OH lines in the raw spectra and masked them to exclude any potential contamination from our final transmission spectra \citep[e.g.][]{Guilluy2023,Guilluy2024,Darpa2024}.\\
        The next step was to isolate the planetary signal via transmission spectroscopy. We shifted all spectra to the stellar rest frame and normalised them to unity. We then computed a master out-of-transit spectrum, $S_\mathrm{out}(\lambda)$, by averaging the out-of-transit spectra with S/N $> 30$. To avoid potential early absorption signals before first contact ($t_1$), particularly evident in \nuno\ (Fig.~\ref{HE_maps}), we restricted the pre-transit baselines to orbital phases $< -0.009$ for \nuno\ and $< -0.0065$ for \nquattro.
        
\begin{figure}
                \centering
                \includegraphics[width=0.9\linewidth,height=10cm]{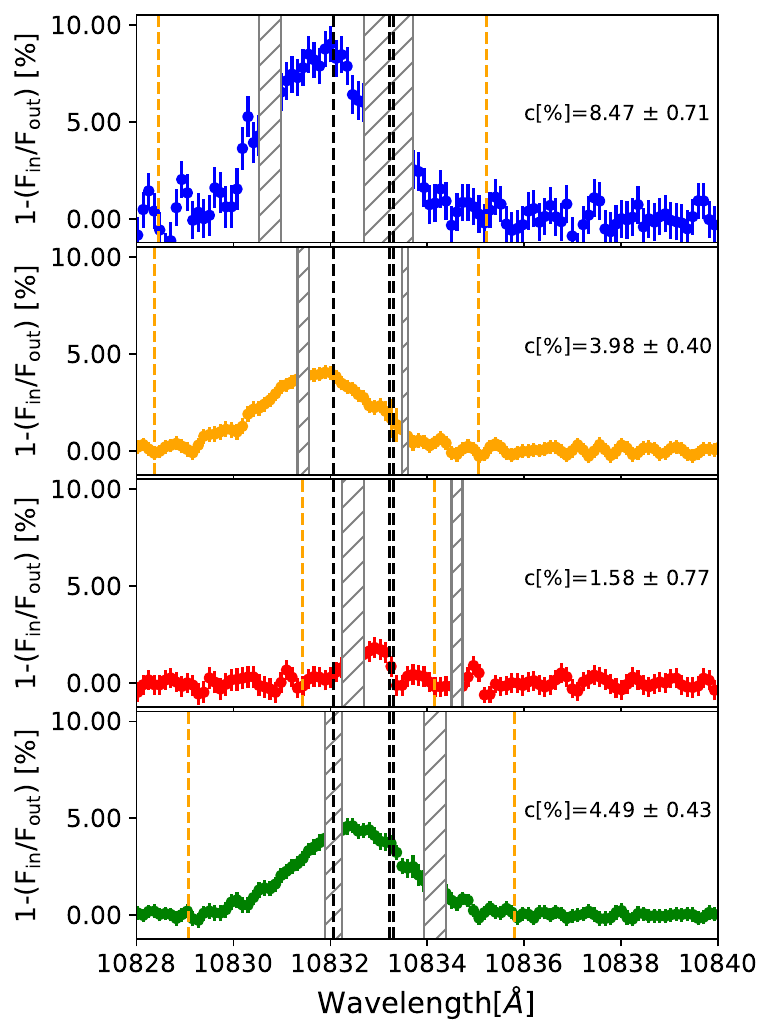}
                \caption{Nightly averaged between contact points $t_2$ and $t_3$, GP-corrected transmission spectra (from top to bottom: \nuno, \ndue, \ntre, \nquattro) with their corresponding absorption depths ($c$ [\%]). Vertical dotted lines indicate the \Hei\ rest-frame positions. Wavelength intervals indicating $\pm$3$\sigma_{\text{Gauss}}$  from the line peak are enclosed by vertical orange lines. Grey shaded areas denote regions masked due to telluric OH doublets.}
                \label{additional}
        \end{figure}

        Individual transmission spectra, $T(\lambda,i)$, were obtained by dividing
        each spectrum by $S_{\rm out}(\lambda)$ and shifting it to the planetary rest
        frame (using the parameters from Table~\ref{tab:master_parameters}). The
        resulting 2D transmission maps are shown in Fig.~\ref{HE_maps}. 
        For each night, we also computed a weighted-average
        in-transit transmission spectrum between contact points $t_2$ and $t_3$ (Fig.~\ref{additional}). We modelled these average profiles with a
        DEMCMC algorithm \citep{TerBraak2006}, fitting a Gaussian profile (peak
        position, full width at half maximum, absorption depth $c$, and continuum
        offset). Correlated noise was mitigated through GP regression with a
        squared-exponential kernel \citep[e.g.][]{Guilluy2024, Darpa2024}, while a
        jitter term $\sigma_\mathrm{j}$ accounted for uncorrelated noise. Using the publicly available \parvati\footnote{\url{https://pypi.org/project/parvati}} tool (Rainer in prep.), we extracted the equivalent widths (EWs) of the GP-corrected weighted-average in-transit transmission spectra within a $\pm 3\,\sigma_{\text{gauss}}$ range around the line peak, resulting in 211.1\,$\pm$\,4.9, 106.3\,$\pm$\,2.1, 19.2\,$\pm$\,1.7, and 123.9\,$\pm$\,2.1~m\AA\ for \nuno, \ndue, \ntre, and \nquattro, respectively. The main parameters of the Gaussian fit, as well as the EWs for each night, are reported in Table~\ref{result}. 

\begin{figure}
                \centering 
                \includegraphics[width=0.92\linewidth]{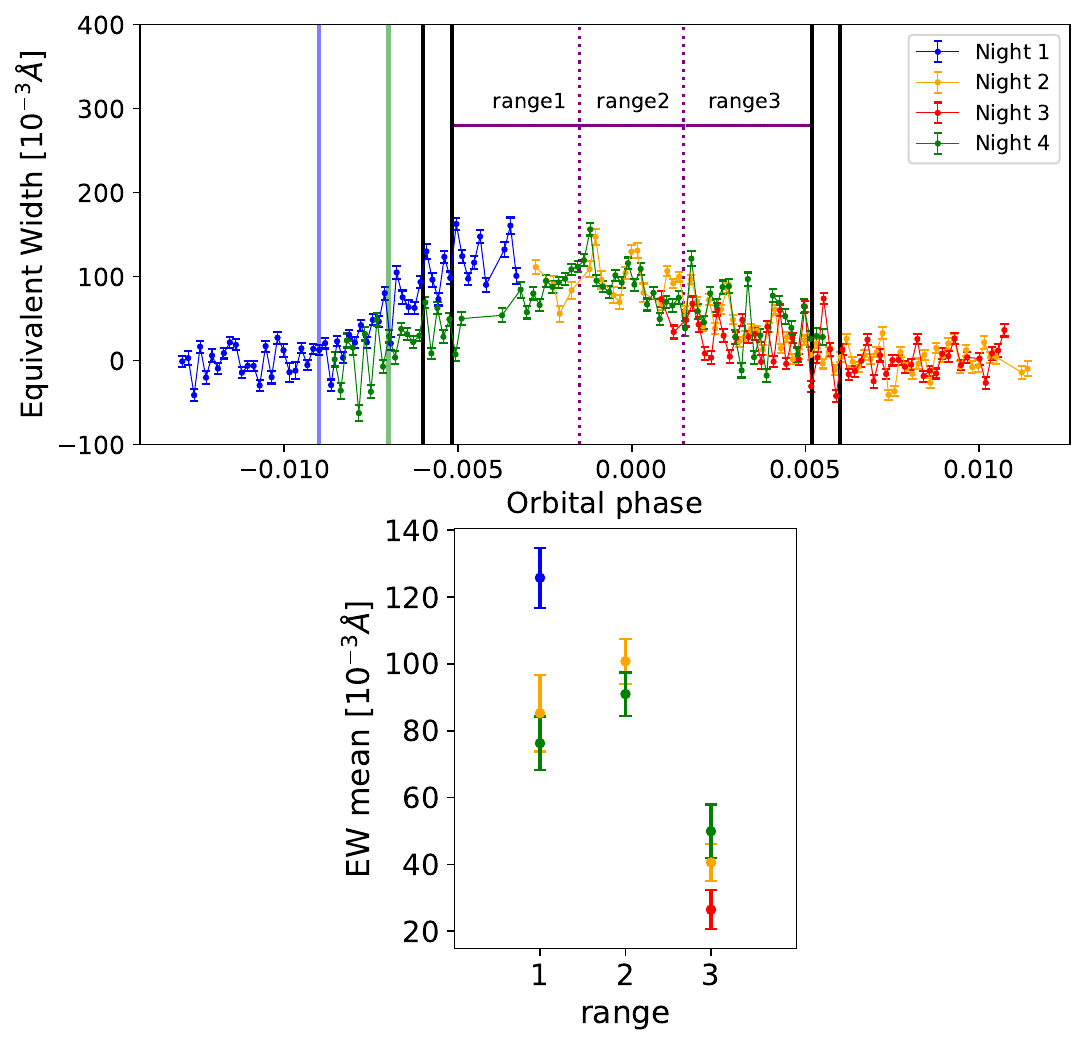}
                \caption{\textit{Top}: EWs of the excess absorption, over the 10830.0–10833.2~\AA\ range in the planetary rest frame. Vertical black lines mark the transit contact points ($t_1$--$t_4$). Blue and green vertical lines indicate the phase thresholds used to define the master out-of-transit spectra for \nuno\ and \nquattro, respectively. \textit{Bottom}: Weighted average EWs computed across three in-transit phase intervals (Ranges 1–3), delineated by vertical purple lines in the top panel.}
                \label{HE_ew}
        \end{figure}

Finally, we calculated the EWs of the individual transmission spectra using a uniform integration passband across all four nights (Fig.~\ref{HE_ew}). Specifically, this window was defined by computing the weighted mean of the Gaussian centroid ($l_0$) and $\sigma_{\text{Gauss}}$ derived from the DEMCMC fits to the nightly mean transmission spectra, setting the integration bounds to $l_0 \pm 1\sigma_\mathrm{gauss}$, that is, 10830.96--10833.14~\AA. We selected a restricted $1\sigma_\mathrm{gauss}$ range (rather than a broader window) to limit contamination from the telluric line at $\sim1083.5\,\text{nm}$ . Since uneven phase coverage across the observing nights can make full-transit averages misleading, we divided the $t_2$--$t_3$ transit chord into three phase intervals: $R_1$ ($-t_2 < \phi < -0.0015$), $R_2$ ($-0.0015 < \phi < 0.0015$), and $R_3$ ($0.0015 < \phi < t_3$). The temporally resolved EWs for each night and phase bin are reported in the bottom panel of Fig.~\ref{HE_ew}.
        The results reported here were obtained using the SM21 mass value. However, they do not change if the mass determination of L26 is adopted, since the derived $K_\mathrm{p}$ for a 24-day orbital period depends only weakly on the planetary mass.
        Given that our measured $T_{14}$ values for both \nuno\ and \ndue\ (Table~\ref{tab:master_parameters}) are higher than literature estimates \citep{David2019a,David2019b,Feinstein2021,Suarez2021,Livingstonetal26}, though still compatible within $1.4\,\sigma$, we tested the robustness of our analysis by shifting the transit center to $T_0 \pm 1\sigma$. The resulting contrast ($c\%$) and EWs of the final averaged transmission spectra remain compatible with our findings.
        
        \subsection{The puzzle of the \texorpdfstring{\Hei}{He I} signal} \label{sec_puzzle}
        Figure~\ref{HE_ew} illustrates the time series of the excess-absorption EWs derived from the GIANO-B transmission spectra around the \Hei\ triplet. The EWs exhibit an asymmetric shape, characterised by stronger absorption before orbital phase $\sim$\,0 than after. This asymmetry, corroborated by the 2D maps in Fig.~\ref{HE_maps}, suggests the presence of a helium structure preceding the planet along its orbit.
        Furthermore, the 2D maps and Fig.~\ref{additional} reveal a noticeable blueshift. However, this shift is not uniform across all observations: \nuno\ appears to be in good agreement with \ndue, while \ntre\  and \nquattro\ exhibit a less pronounced displacement. Notably, \ntre\ also displays a smaller FWHM and depth $c$ than the other nights (Table~\ref{result}).
        \begin{table*} [ht!]
                \centering      
                \caption{Best-fit parameters from the DEMCMC Gaussian fit to the nightly mean in-transit transmission spectra at the \Hei\ triplet position.}
                \label{result}
                
                \begin{tabular}{ccccccc}
                        \hline \hline
                        Night &      $l_0$ &  $v$  & $c$ &         $\sigma_{\text{Gauss}}$& EW(\Hei) \\
                        &     [$\AA$]&   [km/s] &  [\%] &      [nm] &  [m$\AA$]  \\ 
                        \hline
                        1    & 10831.84$^{+ 0.12 }_{ -0.11 }$ & -39.5 $^{+3.2}_{ -3.1 }$ & 8.47 $^{+ 0.72 }_{ -0.71 }$ & 0.1127 $^{+ 0.013 }_{ -0.011 }$    & 211.1$\pm$4.9 \\
                        2     &  10831.72$^{+ 0.12 }_{ -0.12 }$ & -42.7 $^{+3.3}_{ -3.4 }$ & 3.98$^{+ 0.40 }_{ -0.39 }$ & 0.1114 $^{+ 0.016 }_{ -0.013 }$    & 106.3$\pm$2.1  \\
                        3      & 10832.78$^{+ 0.17 }_{ -0.33 }$ & -13.2 $^{+4.8}_{ -9.2 }$ & 1.58$^{+ 0.73 }_{ -0.81 }$ & 0.0453 $^{+ 0.048 }_{ -0.018 }$   & 19.2$\pm$1.7  \\
                        4 & 10832.43$^{+ 0.12}_{ -0.12 }$ & -23.0 $^{+ 3.2 }_{ -3.2 }$ & 4.49$^{+ 0.43 }_{ -0.42 }$ & 0.1122 $^{+ 0.015 }_{ -0.014 }$   & 123.9$\pm$2.1\\
                        \hline
                \end{tabular}
                \tablefoot{From left to right: peak position $\lambda_0$, velocity shift $v$, absorption depth $c$, Gaussian standard deviation $\sigma_{\text{Gauss}}$, and EWs derived within $\pm 3\,\sigma_{\text{gauss}}$ from the peak.}

        \end{table*}
        Several caveats must be considered in this analysis. First, the averaged transmission spectra were obtained over different time intervals owing to the uneven phase coverage of the observing nights. Second, the data are affected by telluric contamination from OH emission lines and $\text{H}_2\text{O}$ absorption. Although the OH lines were masked and the water vapour line was corrected using \texttt{molecfit}, some residuals may persist in the final spectra.  A further complication comes from the fact that the three \Hei\ triplet lines are blended with one another and with the \Sii\ photospheric line (Fig.~\ref{so_out}).
        
        In-transit EW variability among the different nights is also visible. The maximum variability is observed in $\text{range}~1$ between \nquattro\ and \nuno\ at 4.1$\sigma$. Meanwhile, in $\text{range}~2$, the largest variability is observed between \ndue\ and \nquattro, and in $\text{range}~3$, between \ntre\ and \nquattro\,, but the nights remain consistent within $3\sigma$. The variability in $\text{range}~1$ between \nuno\ and \nquattro\ is also visible during ingress. \\
        \indent{Given} the high activity level of \starr, we investigated whether this could drive the observed variability. To this end, we examined the stellar \Hei\ line (Sect.~\ref{sec_stellar_helium}) alongside the \logrhk, \Ha, and Na~\textsc{i} activity indicators \citep[e.g.][]{Guilluy2020} extracted from the HARPS-N spectra (Sect.~\ref{sec_halpha}).
        
        \subsection{Investigation of the stellar helium line} \label{sec_stellar_helium}
        The master-out spectra exhibit significant night-to-night variability at the \Hei\ triplet (Fig.~\ref{so_out}, top panel). Using \parvati, we calculated the stellar  EWs(\Hei) by integrating the flux over the 10831.5--10835.0~\AA\ range. This range differs from the 10830.96--10833.14~\AA\ interval used for the individual planetary transmission EWs (Sect.~\ref{sec_tra}). The stellar integration range was chosen to capture the \Hei\ core while avoiding nearby $\text{H}_2\text{O}$ and blended \Sii\ lines, whereas the planetary range was shifted to capture the absorption feature blueshifted relative to the helium triplet in the 2D maps (Fig.~\ref{HE_maps}). The resulting EWs are 265.0$\pm$1.5, 219.2$\pm$1.2, 172.9$\pm$1.4, and 269.2$\pm$2.7~m\AA\ for \nuno, \ndue, \ntre, and \nquattro, respectively, yielding a maximum variation of $\sim$96~m\AA.  This high variability is also evident in the EWs of the individual spectra (Fig.~\ref{EW_star}, top panel). To provide context, we monitored the nearby photospheric \Sii\ line at $\sim$10830.1~\AA\ (Fig.~\ref{EW_star}, bottom panel). While the \Sii\ line remains remarkably stable across all four nights, the \Hei\ triplet fluctuates significantly. The difference between the \Hei\ and \Sii\ EWs (Fig.~\ref{EW_star}, bottom panel) reveals an in-transit excess, raising the question of whether this signal stems from an evaporating planetary atmosphere.
        
        However, any in-transit signal must be interpreted with caution, given the well-documented intrinsic variability of \starr\ \citep{vissapragada_search_2021, gaidos_zodiacal_2022, Krolikowski2024, alam_non-detections_2024}. Previous studies have reported severe stellar \Hei\ fluctuations: $\sim$0.2~\AA\ over a two-month baseline \citep{vissapragada_search_2021}, variations of up to 50\% during a single transit (likely due to the extended gas tail of planet d, \citealt{gaidos_zodiacal_2022}), and short-term changes of $\sim$80~m\AA\ over just three weeks \citep{Krolikowski2024}. Due to this variability, \citet{Krolikowski2024} suggested a highly conservative 95\textsuperscript{th} percentile detection limit of $\sim$6.7\%. 
        
        Ultimately, stellar activity remains a critical bottleneck that can easily mimic or mask planetary \Hei\ absorption \citep[e.g.][]{Salz2018, Cauley2018, Guilluy2020}. Our findings reinforce that, for active young stars such as \starr, acquiring contiguous out-of-transit spectra immediately before and after the transit is essential to robustly isolate planetary signals \citep{Vissapragada2020, Krolikowski2024}. This stands in stark contrast to older, inactive systems, where multi-epoch baselines are usually sufficient \citep[e.g.][]{Gully2023, Orell-Miquel2025}.

        \subsection{Stellar activity monitoring in the visible} \label{sec_halpha}
        \begin{figure}
                \centering
                \includegraphics[width=0.78\linewidth]{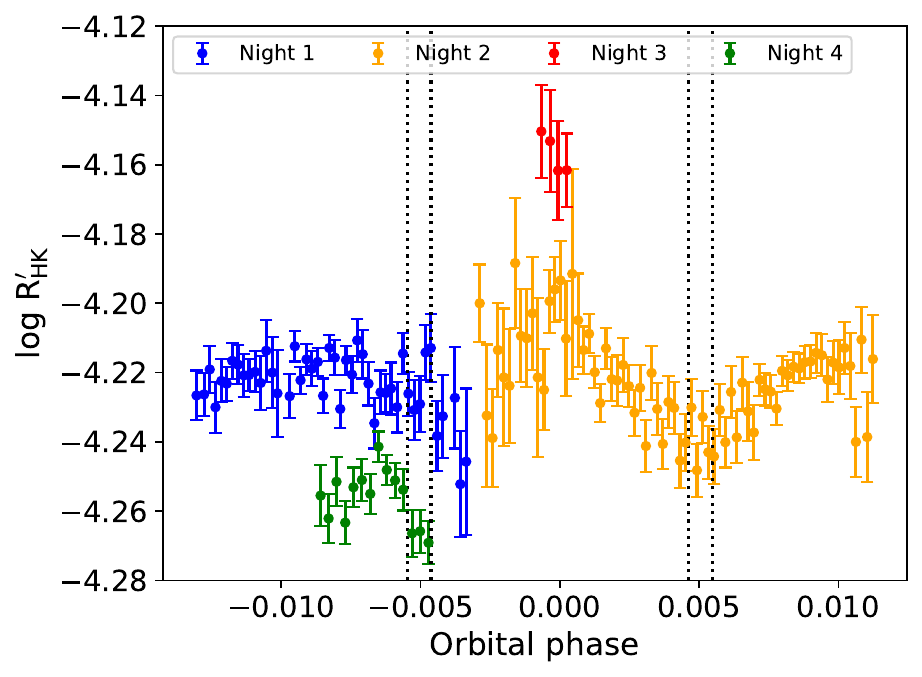}
                \caption{\logrhk\ activity index as a function orbital phase.}
                \label{logrhk}
        \end{figure}

        \subsubsection{\texorpdfstring{\logrhk}{log R'HK}}
        
        We first examined the \logrhk\  chromospheric activity index using HARPS-N data. The index was extracted from the spectra via the YABI platform \citep{yabi}, hosted at the IA2 Data Center\footnote{\url{https://www.ia2.inaf.it/}}, assuming a $B-V$ colour index of $0.99~\text{mag}$. The resulting time series are shown in Fig.~\ref{logrhk}. As detailed in Sect.~\ref{sec_observations}, instrumental issues during \ntre\ and \nquattro\ limited the datasets to only a few pre-transit and ingress spectra and a small number of in-transit spectra, respectively. Conversely, \ndue\ features a detectable rise in the chromospheric index, which we attribute to surface inhomogeneities. This interpretation is corroborated by simultaneous photometric observations that we performed with the Nordic Optical Telescope (NOT), which revealed a strong anomaly on the light curve that is attributable to a large starspot (complex) on the surface of the parent star (Appendix~\ref{not}). This perfectly aligns with the cross-correlation function (CCF) analysis by \citet{maio2023}, who mapped a persistent circumpolar spot visible throughout the rotation, alongside rotationally modulated, low-latitude active regions. Such a scenario is further supported by TESS light-curve analyses (private communication). Recently, \citet{kronos2} found that \starr\ has starspots with distinct umbral and penumbral structures with thermal and geometrical properties similar to those of the Sun. 
        
        \subsubsection{\texorpdfstring{\Ha}{H-alpha}}
        To assess whether the variability observed in the \Hei\ absorption is driven by stellar activity, we analyzed the \Ha\ chromospheric line at 6562.8~\AA\ collected simultaneously to our near-infrared observations as already done in \cite{Guilluy2024} and in \cite{Darpa2024}. While the \Hei\ triplet is primarily populated in the upper chromosphere and transition region via XUV irradiation, the \Ha\ line originates in the lower-to-middle chromosphere through the $n = 3 \to 2$ transition. 
        As a result, \Ha\ serves as a robust tracer of chromospheric features, such as active regions (plages), filaments, and flares. Unlike other young stars of similar activity, which typically display \Ha\ in emission, \citet{Feinstein2021} noted that \starr\ shows a purely absorption profile without a significant core component. 
        Following the strategy adopted for \Hei,  we analysed the master out-of-transit spectra, which revealed a large variability (bottom panel of Fig.~\ref{so_out}), with EWs(\Ha) values of 279.2\,$\pm$\,1.4~m\AA, 270.6\,$\pm$\,1.4~m\AA, and 329.4\,$\pm$\,2.4~m\AA, for \nuno, \ndue, and \nquattro, respectively.  Left and middle panels of Fig.~\ref{logrhkew} show how the EWs of the \Hei\ (left panel) and \Ha\ (right panel) stellar lines (derived from the master-out spectra) vary as a function of \logrhk.   While the \Ha\ line exhibits solar-like behaviour characterised by weaker absorption in more active regimes due to chromospheric filling-in \citep{Cauley2016, Cauley2017}, the \Hei\ triplet show a departure from this pattern. 
        In the Sun, \Hei\ absorption typically intensifies in active regions, with the EW increasing alongside stellar activity. In the solar case, \Ha\ and \Hei\ become typically correlated during relatively strong flares, often preceded by a marked dimming \citep{Kerr2021}.
        In the solar chromosphere, the \Hei\ triplet forms via absorption from the $2^3\text{S}$ atomic level, which is populated through the recombination of photoionised helium atoms (the photoionization-recombination $\text{PR}$ mechanism). Ionising helium requires photons with wavelengths shorter than $503~\text{\AA}$, originating from strong emission in the transition region and corona of active regions. As the activity level increases, this high-energy emission grows, increasing the population of metastable helium and thus resulting in a deeper absorption line and a larger EW.
        For active dwarf stars, where the chromosphere is hot and dense enough, there is an alternative mechanism that contribute to PR to populate the \Hei\ state directly from the ground state via collision (the collisional excitation, $\text{CE}$, mechanism, \citealt{Sanzforcada2008}). In dwarf stars, the increasing levels of activity do not cause more substantial \Hei\ absorption. On the contrary, the absorption appears rather shallow (if not partly in emission) for very active stars.
        In \starr, we observe a decrease in EW(\Hei) with increasing activity, suggesting that the CE mechanism is also contributing to the population of the \Hei~$2^3\text{S}$ level. However, since the averaged \logrhk\ values in Fig.~\ref{logrhkew} span different temporal windows (Fig.~\ref{logrhk}), this trend warrants caution.
        \\
        \begin{figure*}
                \centering
                \includegraphics[width=0.5\linewidth]{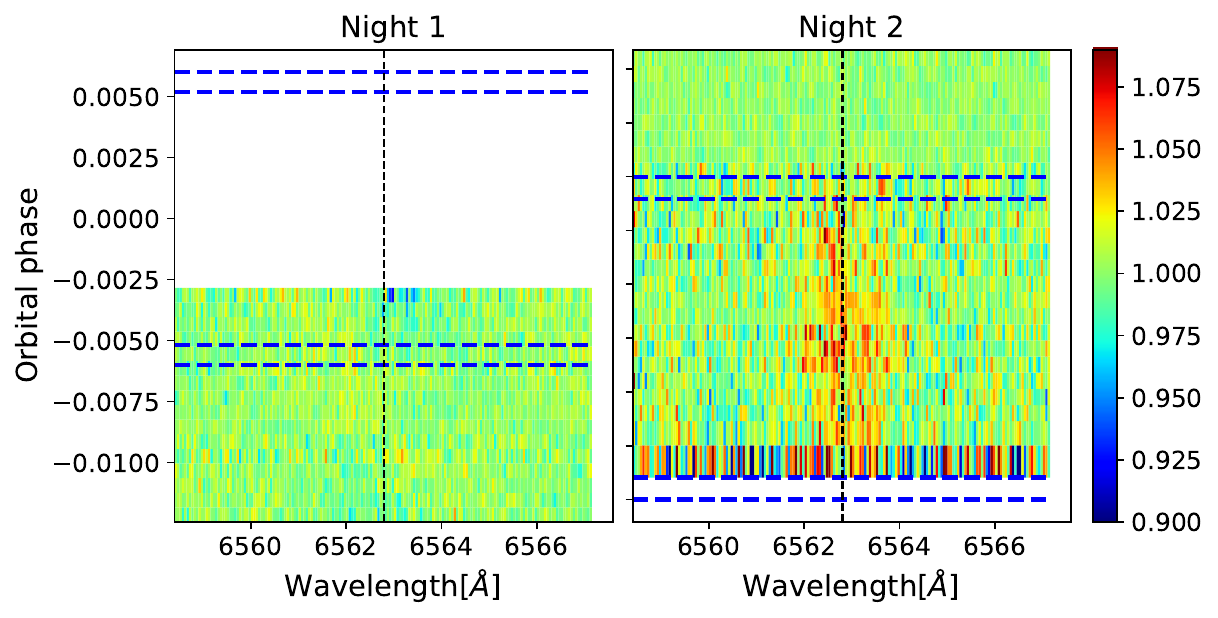}
                \includegraphics[width=0.45\linewidth]{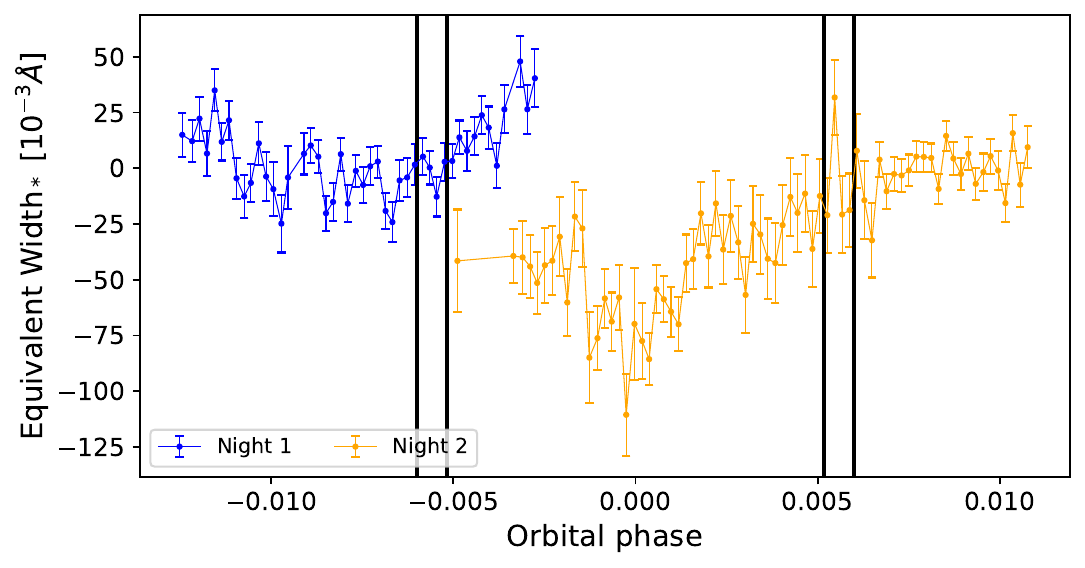}
                \caption{\textit{Left}: 2D maps of the \Ha\ transmission spectra binned in phase and in wavelength, aligned in the planetary rest frame. \textit{Right}: Time-series of the EWs derived from the transmission spectra using the \parvati\ routine, integrated over the 6561.4\,--\,6564.2~\AA, spectral range. A negative EW indicates an emission signal. Since we are measuring the EW of transmission spectra, where lines are typically expected in absorption, a decrease in flux below the continuum yields a positive EW, while an increase above the continuum yields a negative EW. \ntre\ and \nquattro\ were excluded from this specific analysis due to the insufficient number of in-transit spectra.}
                \label{tra_sp_halpha}
        \end{figure*}
        \indent{Following} a similar methodology applied to the \Hei\ triplet, we then derived  \Ha\ transmission spectra. We extracted them following the framework described in detail in \cite{darpa_kelt9}.
        The resulting transmission spectra 2D maps are displayed in the left panel of Fig.~\ref{tra_sp_halpha}. while the right panel illustrates the time-series EWs of these transmission spectra. Notably, \ndue\ exhibits a reduced EW during the transit, accompanied by a distinct emission-like feature in the 2D tomography maps. This signal is most likely attributable to stellar activity. This pseudo-emission feature, which manifests as a relative decrease in the stellar \Ha\ absorption depth, could stem from either a stellar flare occurring during the transit (inducing chromospheric filling-in) or the occultation of a chromospheric active region. Simultaneous NOT photometry strongly supports the latter scenario (Appendix~\ref{not}), indicating that the planet crossed a large starspot (complex) during \ndue. In a standard framework, crossing a chromospheric active region (where the \Ha\ line is typically shallower due to localized emission filling-in) is expected to remove a lower-absorption component from the integrated disk, thereby resulting in a net absorption anomaly in the transmission spectrum (an increase in the observed absorption depth). However, our observations reveal the opposite behavior (i.e. a net pseudo-emission signature). To reconcile this, we hypothesize a geometric configuration where the chromospheric active region (plage) is significantly more extended than both the planetary disk and the underlying photospheric spot. This is consistent with solar observations, where chromospheric faculae generally encompass much larger areas than their photospheric counterparts. In this framework, the specific chord crossed by the planet may be locally less active (characterized by a lower emission contrast) than the surrounding, unocculted sections of the plage. Consequently, as the planet blocks this quieter zone, the emission from the dominant, unocculted active region decreases the net integrated \Ha\ absorption depth of the star during the transit.
        Regarding the \Hei\ signal during \ndue, we observe the opposite trend: an increased absorption (a deeper line profile) at mid-transit (Figs.~\ref{HE_maps} and~\ref{HE_ew}). As discussed in Sect.~\ref{sec_halpha}, the stellar \Hei\ absorption depth in \starr\ decreases with increasing stellar activity due to chromospheric filling-in. Therefore, a spot-crossing event involving an active region with a shallower \Hei\ profile results in a net absorption peak in the in- and out-of-transit ratio, as the `less deep' component is temporarily blocked by the planet. 
        The simultaneous decrease in \Ha\ absorption depth and the increase in \Hei\ absorption can be physically reconciled by assuming that the formation regions of these two indicators are spatially decoupled (e.g. forming at different chromospheric heights). This atmospheric stratification allows the planet to simultaneously occult a region that appears less active in \Ha\ relative to its immediate surroundings, while remaining sufficiently active in the \Hei-forming layers to generate the observed absorption enhancement. We note that a remarkably similar behavior, an increase in \Hei\ absorption coupled with an \Ha\ emission anomaly, has been reported by \citet{Orell-Miquel2023} for the 650-Myr-old sub-Neptune HD~235088 b.
        Yet, these considerations are qualitative; further observations providing full transit coverage, including robust pre- and post-transit baselines, are essential to disentangle the planetary signal from the complex stellar activity of this system. In contrast, the \Ha\ transmission spectroscopy for \nuno\ reveals a marginal absorption signal. Due to the low $K_\mathrm{p}$ value (Table~\ref{tab:master_parameters}), which is a consequence of the planet's relatively long orbital period, it is challenging to discern whether this feature is better aligned with the planetary or the stellar rest frame. However, the fact that \ndue\ displays a pronounced emission signal suggests that the features observed on both nights are dominated by stellar activity, and reflect the complex stellar surface.
        
        \subsubsection{\ion{Na}{i}}
        In addition to \Ha\ and \logrhk, we also analysed the \ion{Na}{i} doublet. Our analysis reveals a significant night-to-night variability. We observe a clear emission signal in \ndue. Conversely, \nuno\ is affected by significantly larger uncertainties that preclude a definitive conclusion. A comprehensive discussion about the analysis is reported in Appendix~\ref{sodium_app}.
        
        \section{Discussion} \label{Discussion}
        \subsection{Different stellar activity levels can lead to different \Hei\ planetary absorption signals} \label{ATES}
        An additional complication in separating the stellar contamination from the planetary signal arises from the fact that the observed planetary absorption may be altered by variations in the stellar flux caused by activity.
        To test how different activity levels affect the helium absorption profile, we simulated the evaporation of \planet\ with the hydrodynamic code ATES \citep{Caldiroli+2021}, using different XUV irradiation levels. We scaled the reference XUV spectral energy distribution (SED, \citealt{maggio2023}) by factors of 0.5, 1, and 2 relative to its baseline integrated flux $F_0 = 3.73 \times 10^4$ erg~s$^{-1}$~cm$^{-2}$. The simulations were performed for both the tabulated mass values (Table \ref{tab:master_parameters}) fixing the H/He ratio at 100/10; the results are presented in Table \ref{tab_ates}.
        Although core-powered mass loss is not accounted for in ATES and could complicate the atmospheric escape picture, we expect it to be subdominant over the range of planetary masses and equilibrium temperatures considered for \planet\ \citep{Owen2024,Misener2026}.

        In the case of the higher planetary mass value from SM21, the \textsc{ATES} simulations indicate that increasing the \text{XUV} flux leads to a decrease in the predicted transit depth. Since the outflow is highly ionized, a higher number of ionizing photons lowers the density of singly ionized helium -- which would otherwise supply the metastable state via recombination -- in the diffuse, optically thin region of the outflow, which dominates the absorption signal (Fig.~\ref{fig:ATESHeTR}, bottom panel). Accordingly, the predicted depth and EW of the metastable helium line decrease as the XUV flux increases. 
        The opposite trend is found for the lower-mass L26 case: increasing the XUV flux produces a stronger outflow that remains dominated by neutral and singly ionized species. As a result, the population of metastable helium grows with irradiation (Fig.~\ref{fig:ATESHeTR}, top panel), leading to a larger predicted line depth and EW.
        
        It is important to note that reproducing the exact absolute values of the observed \Hei\ depth and EW remains a general challenge for 1D hydrodynamic frameworks. As recently discussed by \citet{Taylor2026}, 1D models frequently underestimate the total observed EW because they omit additional non-thermal broadening mechanisms. Processes such as 3D atmospheric dynamics, plasma turbulence, or magnetic interactions with an active host star can significantly broaden the absorption line profile. This extra velocity dispersion yields observed EWs that are systematically larger than those predicted by purely radial outflow models.
        The observed depths for \ndue, \ntre, and \nquattro, plotted as a function of \logrhk\ (last panel of Fig.~\ref{logrhkew}), appear to follow the trend predicted by the simulations when assuming SM21 mass: higher stellar activity leads to a greater $\text{XUV}$ flux and, consequently, a reduced \Hei\ absorption depth. The only clear exception is \nuno. 
        The discrepancy in \nuno\ can be attributed to two factors: an enhanced star-planet interaction, suggested by the leading tail observed exclusively on this night (Sect.~\ref{sec_star_planet}), and potential residual telluric contamination, given that the 10835.1~\AA\ telluric line completely overlaps with the stellar \Hei\ line during this particular run (Sect.~\ref{sec_puzzle}; Fig.~\ref{rest_frame}).
        It is worth noting, however, that because the measured absorption depths are derived from averages over different temporal windows (e.g. \ntre\ covers only the final part of the transit), this trend should be interpreted with caution.

        \subsection{Comparison with previous works}
        Previous analyses of the \Hei\ triplet in \planet\ resulted in non-detections from partial transits \citep{vissapragada_search_2021} or ambiguous results \citep{gaidos_zodiacal_2022}. Specifically, \cite{gaidos_zodiacal_2022} analyzed multiple observations of \planet\ across several nights in 2020, covering the transit along with pre- and post-transit baselines. To minimize baseline systematics and stellar variability, we construct transmission spectra exclusively using out-of-transit references obtained on the same night as the transit, whereas \cite{gaidos_zodiacal_2022} used out-of-transit spectra obtained on the preceding night. In their study, they found an elevated absorption during the transit relative to the pre-transit baseline; however, the absorption decreased throughout the transit, with an EW variation of $\Delta\text{EW}(\text{\Hei}) \approx 0.19\,\text{nm}$. 
        \cite{gaidos_zodiacal_2022} noted that this behavior could be explained by a He-bearing cloud preceding \planet\ in its orbit. Their observation sequence began only 1\,h after the transit of planet~d, introducing potential contamination from this interior planet. We avoided this issue by selecting observation windows free from orbital phase overlap (Sect.~\ref{sec_observations}). 
        While a direct quantitative comparison between our EW(\Hei) values and those of \cite{gaidos_zodiacal_2022} is not feasible due to differences in calculation methodology, we can compare the shape of the absorption. Notably, the feature shown in their Fig.~8 appears fully compatible with both the wavelength amplitude and the blueshift relative to the \Hei\ rest wavelengths observed in our dataset (Fig.~\ref{HE_maps}).
        
        \subsection{The role of star-planet interactions in shaping the \Hei\ transit signature} \label{sec_star_planet}
        In the previous sections, we highlighted the difficulty in disentangling activity-induced pseudo-signals from real planetary absorption for a highly active star such as \starr. We also demonstrated how the expected metastable \Hei\ absorption is strongly modulated by the incident stellar flux, which governs the population of the $2^3S$ state. These combined factors could potentially contribute to the diverse absorption levels observed across the different investigated epochs. Furthermore, \nuno\ exhibits a distinct blueshifted leading tail, a feature that is counterintuitive based on simple orbital kinematics and the conservation of angular momentum, which would typically predict a redshifted signal in a trailing wake. Crucially, this blueshifted signature is only visible during \nuno\ and not during \nquattro. According to the \logrhk, \nuno\ shows a higher activity level than \nquattro. This may suggest that the anomalous velocity component responsible for the blueshift is not an intrinsic feature of the planetary outflow under quiescent conditions, but is instead directly modulated by variations in the stellar environment (XUV flux, wind properties, and magnetic configuration) and the local plasma regime governing the star-planet interaction. Additional insight is provided by the averaged in-transit transmission spectra (Fig.~\ref{additional}), where night 1, displaying the clearest pre-ingress absorption, is also associated with the strongest and broadest He I. This combination disfavours a scenario in which the leading feature is produced by a purely static, magnetically trapped reservoir alone. Instead, it suggests that the absorbing material has a significant dynamical component (contributing to both the line broadening and the net Doppler shift, Table~\ref{result}), while still being shaped and, at times, concentrated by confinement within the planet’s environment .
        
        The nature of the star-planet interaction depends on the local Alfvénic Mach number $M_{\rm A}=u/v_{\rm A}$, defined as the ratio of the stellar wind velocity $u$ to the local Alfvén speed, which is defined as $v_A=B/\sqrt{4\pi\rho}\,$, where $B$ and $\rho$ are the magnetic field strength and plasma density of the wind at the planetary orbit. When this ratio exceeds unity (super-Alfvénic regime), a bow shock can form; when it is below unity (sub-Alfvénic regime), no shock develops. The Alfvénic Mach number at the planetary orbit depends on stellar wind parameters, such as the stellar mass-loss rate and surface magnetic field strength, that remain poorly constrained, making it difficult to determine unambiguously which regime applies. Moreover, variations in stellar activity level may cause the system to alternate between the two regimes because such variations are accompanied by changes in the intensity and in the configuration of the stellar field close to the planet.
        
        We conjecture that V~1298~Tau~b is located in a region where the Alfvenic Mach number of the stellar wind is of the order of unity ($M_{\rm A} \sim 1$) for most of the time. In such a regime, assuming a planetary dipole field of $B_{\rm p} \sim 25$~G and a stellar dipole field of $B_{\rm s} \sim 200$~G \citep{Finocietyetal23}, the extension of the planetary magnetosphere in the direction of the impinging stellar wind is of $\sim 12\, R_{\rm p}$, where $R_{\rm p}$ is the radius of the planet (Appendix \ref{appendix_wind}). For a planet mass of $\sim0.042$~\mjup\ (L26), the low surface gravity of the planet and the heating of the planetary  atmosphere by the strong stellar XUV flux make its evaporation remarkable, thus filling all the magnetosphere with plasma. The large pressure scale height of the plasma confined inside the magnetosphere,  $H_{\rm p} \sim 1.6\, R_{\rm p}$, makes the density difference between the surface of the planet and the boundary of the magnetosphere less than a factor of two for a magnetospheric plasma temperature of $T =10^{4}$~K (Appendix~\ref{plasma_distr}). 
        
        To explain why the leading tail of helium is observed only during \nuno, we can conjecture that, only during that night, the magnetosphere of the planet was remarkably compressed by the interaction with a faster and denser stellar wind streamer with an Alfvenic Mach number $M_{\rm A} \sim 10$. The compression of the magnetosphere by a factor of about two in the direction of the streamer made its internal density increase by approximately the same factor up to a distance of $\sim 6\, R_{\rm p}$ in the direction of the streamer. During \nuno, the master-out spectrum was taken before the transit and before the compressed magnetosphere was transiting over the stellar disc. Later,  the compressed magnetosphere started to transit over the stellar disc, thus providing an excess absorption with respect to the master-out spectrum. This corresponds to the detection of a leading helium tail before the first contact during \nuno. The angle $\theta_{0}$ formed by the impinging wind streamer with the direction tangent to the planetary orbit depend on the properties of the stellar wind, specifically, its velocity components, and the planet orbital velocity \citep{Llama2013}. It controls the duration of the helium absorption in the leading tail by projecting the radius of the compressed magnetosphere by a $\cos \theta_{0}$ factor along the orbit of the planet. 
        
        During \nquattro, the leading tail was not observed. We interpret such a result as the stellar wind returning to its quiescent regime ($M_{\rm A} \sim 1$) leading to a magnetosphere with a lower density and a larger extension in the direction of the wind with only a weak shock, if any, at the boundary. On that night, the master-out corresponded to such a lower density in the extended magnetosphere. Its inner part was denser by a factor of about two than its boundary (cf. Appendix~\ref{plasma_distr}), thus explaining the absorption detected in transit during \nquattro. 
        
        Finally, during \ndue ~and \ntre, only the second part of the transit was observed. This corresponded to the transit of the trailing planetary magnetosphere that was not remarkably impacted by the stellar wind, assumed to be in its quiescent regime, and the density of which was higher close to the planet. On those nights, we saw no significant density differences out of transit when using a master-out obtained after the fourth contact because the density in the extended magnetosphere behind the planet was varying slowly due to the large pressure scale height.
        
        If we adopt a planet mass of $\sim0.63$~\mjup\ (SM21), the density gradient across the magnetosphere becomes unrealistic and our interpretation is no longer valid (cf. Appendix ~\ref{plasma_distr}). Adopting the planet mass reported in L26, our model can also account for the broadening of the neutral helium triplet and its variations from one night to the other (cf. Fig.~\ref{additional}), at least qualitatively. We discuss this further in Appendix~\ref{appendix_wind}.

        \subsection{Planetary evolution} \label{evolution}
        \begin{figure*}[h!]
                \begin{center}
                        \begin{tabular}{ccc} 
                                \includegraphics[width=5.4cm]{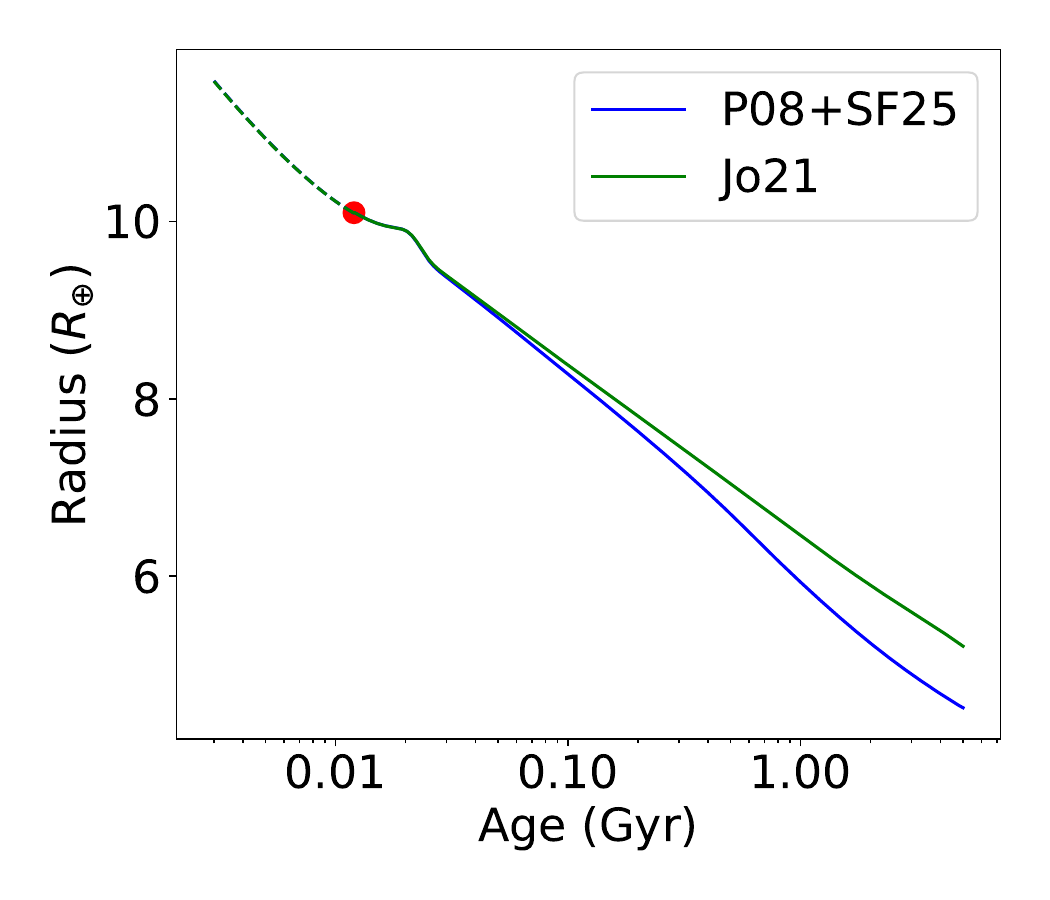} & 
                                \includegraphics[width=5.4cm]{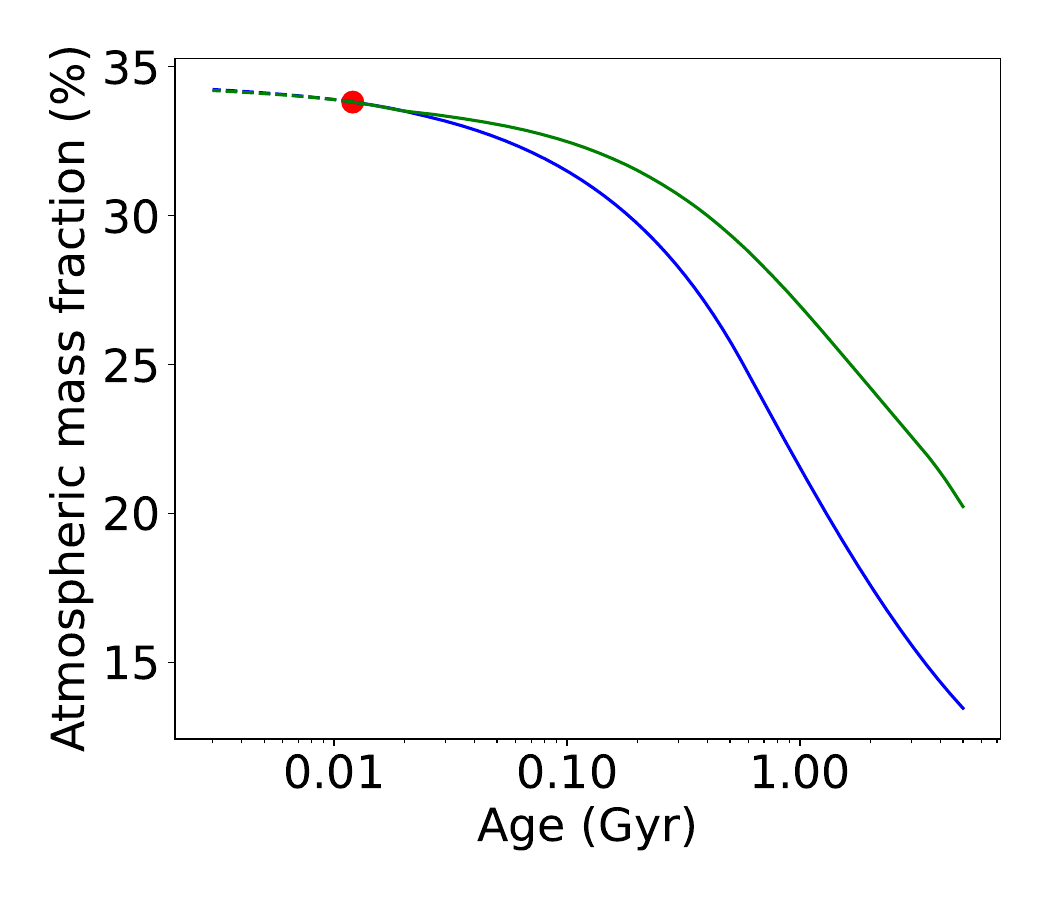} & 
                                \includegraphics[width=5.4cm]{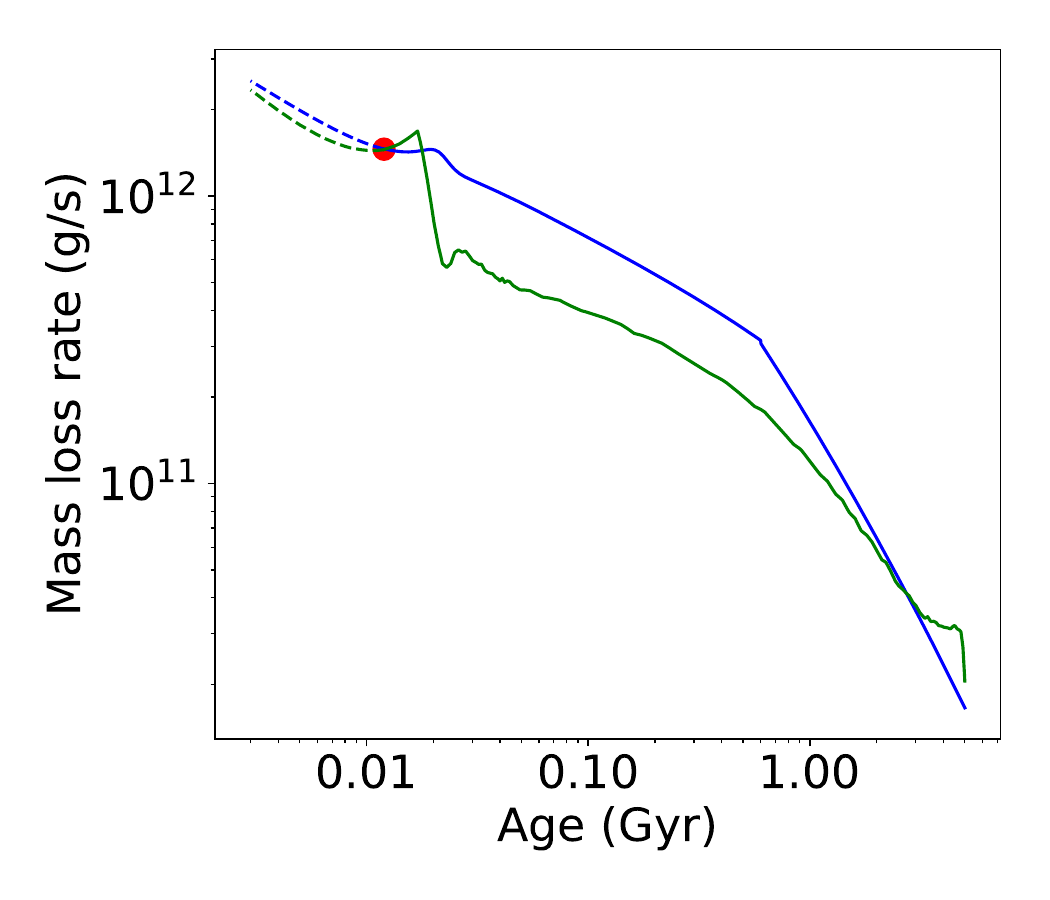} \\
                        \end{tabular}
                        \caption{Evolution of radius  (\textit{left}), atmospheric mass fraction (\textit{middle}), and mass-loss rate (\textit{right}) of V1298\,Tau\,b considering the mass from L26. Solid and dashed lines represent future and past evolution, respectively. The red circle shows the position of \planet\ at 12 Myr.}
                        \label{fig:evap}
                \end{center}
        \end{figure*}
        The hydro-D simulations with ATES provide us with an estimate of the atmospheric mass loss rate at the present age of the system. In this respect, we were able to recompute the stellar age by matching the position of V1298\,Tau in the temperature-luminosity diagram with the MESA evolutionary tracks \citep{Choi+2016}. To this end, we assumed the stellar mass and effective temperature reported in Table~\ref{tab:master_parameters}, 
        and $L_{\rm bol} = 0.943 \pm 0.018 \, L_{\odot}$. These values lead to an age of $12 \pm 1$\,Myr, which is in agreement with the results of \cite{Finocietyetal23}, but significantly younger than the age of $23 \pm 4$\,Myr obtained by \cite{David2019a} using magnetic-inhibited convection models of stellar interiors \citep{Feiden2016}. In the following, we adopt this younger age. We recommend the reading of \cite{gaidos_zodiacal_2022} for a detailed analysis of the uncertainties on the age of this system.
        
      To investigate whether photoevaporation may produce relevant variations of the planetary radius and mass, an educated guess of the long-term evolution of the stellar XUV emission is required.
        Following the approach first developed in \cite{Locci19} and subsequently updated to study the evolution of single systems (e.g. \citealt{maggio_new_2022, Damasso+2024b, Nardiello+2025}), we have simulated the potential photoevaporation history of V1298\,Tau\,b, from an age of 3 Myr to about 5 Gyr. 
        In our modelling, we adopted the ATES code \citep{Caldiroli+2021,Caldiroli2022,Spinelli2023} the planetary core-envelope models by \cite{Fortney+2007} and \cite{Lopez+Fortney2014}, and the MESA Stellar Tracks \citep{Choi+2016}. Initially, we assumed a planetary mass of $\sim0.042$~\mjup\ (L26) and computed the internal structure of the planet, obtaining a core mass of $8.8\,M_\oplus$ and a core radius of $1.7\,R_\oplus$. Moreover, at the current system age, we obtained an atmospheric mass fraction of $33.8\%$.
        
        We then evolved the system forward and backward in time, updating the planetary radius to account for natural gravitational contraction and photoevaporation-driven atmospheric mass loss. For the stellar XUV luminosity history, we perform two simulations: the first is based on \cite{Penz+2008} coupled with the X-ray-to-EUV scaling law by \cite{Sanz-Forcada25} (hereafter P08+SF25), while the second adopts the semi-empirical modeling of \cite{Johnstone2021} (hereafter Jo21). In both cases, we estimate a current mass loss rate of $\sim 1.45 \times 10^{12}$\,g\,s$^{-1}$.

        The predictions of the two simulations at 3 Myr are similar. In both cases, the planet begins its evolution with a mass of approximately $13.4\,M_\oplus$, an atmospheric mass fraction of about $34.2\%$, and a radius of $11.6\,R_\oplus$. Since the P08+SF25 tracks predict a higher XUV luminosity, the corresponding evaporation rates are larger in this scenario, leading to a more substantial mass loss.  At 5 Gyr, the P08+SF25 model predicts a planetary radius of approximately $4.5\,R_\oplus$ and an atmospheric mass fraction of $13.5\%$, whereas the Jo21 model provides a radius of $5.2\,R_\oplus$ and an atmospheric mass fraction of $20.2 \%$. In Fig.~\ref{fig:evap} we show the temporal evolution of the planetary radius, atmospheric mass fraction, and mass-loss rate.
        Both evolutionary scenarios indicate that the planet loses more than half of its primordial atmosphere and its radius contracts evolving into a mature sub-Neptune, in agreement with \cite{Barat2025} and L26. 
        Alternatively, when we performed the simulation by assuming a mass of $\sim0.63$~\mjup\ (SM21), we found that the deeper gravitational potential ensures that \planet\ remains scarcely affected by photoevaporation induced by high-energy radiation, thereby confirming the conclusions of \cite{maggio_new_2022}. Conversely, a direct comparison with the findings of \cite{maggio_new_2022} is not feasible when considering the mass from L26, due to the substantial discrepancy in the assumed mass values.

        \section{Conclusions} \label{sec_conclusion}
        In this work, we investigate the metastable \Hei\ triplet of \planet\ by analysing four transit events observed with the high-resolution GIANO-B spectrograph. Due to the long transit duration ($\sim$7~hours), the dataset exhibits uneven phase coverage and lacks comprehensive out-of-transit baselines, missing contiguous pre- and post-transit observations. This poses significant challenges for characterising a planet orbiting such an active host star. Further complexity arises from the intrinsic blending of the \Hei\ triplet components with the nearby photospheric \Sii\ line, due to the fast rotation of the star, as well as significant telluric contamination from OH emission and $\text{H}_2\text{O}$ absorption. Finally, the low planetary $K_\mathrm{p}$ complicates the disentanglement of the signals, making it difficult to definitively align the observed features with either the stellar or the planetary rest frames.
        
        We first modelled the RML effect, refining the projected spin-orbit angle to $\lambda$ = $31 \pm 14$~deg. While this value suggests a potential departure from exact alignment, the large uncertainties make it fully consistent with previous literature estimates. We then performed \Hei\ transmission spectroscopy, reporting the detection of \Hei\ absorption in all four analyzed nights; however, we observed significant night-to-night variability. The measured absorption depths ($c$) are $8.47\pm0.71\%$, $3.98\pm0.40\%$, $1.58\pm0.77\%$, and $4.49\pm0.43\%$, with corresponding EWs of $154.0\pm4.5$~m\AA, $81.4\pm1.9$~m\AA, $18.8\pm1.7$~m\AA, and $113.4\pm2.0$~m\AA\  for \nuno, \ndue, \ntre, and \nquattro, respectively. We need to emphasise that these values are derived from time-averaged in-transit spectra, which were obtained over different phase intervals due to the uneven coverage of the dataset. Specifically, for \ntre, the observations only cover the final part of the transit.
        
        In agreement with previous studies \citep[e.g.][]{vissapragada_search_2021, gaidos_zodiacal_2022, Krolikowski2024, alam_non-detections_2024}, the master out-of-transit spectra also exhibit variability, which is driven by the strong activity of the young star. To disentangle stellar activity from planetary signals, we also analysed the \logrhk\ and \Ha\ indicators, despite the limited data coverage for \ntre\ and \nquattro. The variability in the master-out EWs of both \Hei\ and \Ha\ against \logrhk\ underscores the significant impact of the active regions on these lines. In the \Ha\ transmission spectra, we observed contrasting behaviours: \ndue\ exhibits a pronounced in-transit pseudo-emission `bump' (likely attributable to the occultation of a large stellar starspot or spot complex, as indicated by NOT simultaneous observations), whereas \nuno\ shows only a marginal absorption signal. Although these \Ha\ features are clearly dominated by stellar activity, which likely contaminates the \Hei\ measurements, a key distinction emerges. While the \Ha\ indicator displays opposite trends across different epochs, a behaviour further supported by the Na~\textsc{i} analysis, which reveals an emission signal during \ndue\ but is affected by significantly larger uncertainties in \nuno; the \Hei\ triplet shows an absorption signal across four independent measurements spanning a four-year baseline, albeit with varying depths. This may support a planetary origin for the \Hei\ signal, although the underlying stellar activity inevitably modulates its observed depth.\\
        \indent{This} interpretation is supported by our \textsc{ATES} simulations, which demonstrate that stellar XUV flux variations significantly alter the absorption depth. We show that the response of the \Hei\ triplet to XUV radiation is highly sensitive to planetary mass, shifting from an over-ionization regime (and thus reduced absorption) for the higher-mass scenario (SM21) to an optically thick outflow dominated by neutral and singly ionized species, resulting in enhanced absorption, for the lower-mass case (L26).\\
        \indent{We} attribute the blueshifted leading tail, seen only during \nuno, to a compression of the planetary magnetosphere likely triggered by a fast and dense stellar wind streamer. This explains why the tail vanished during the less active \nquattro, despite the pre-transit coverage that would have allowed its detection.\\
        \indent{Finally}, we investigated the evolution of \planet\ to assess whether photoevaporation induces significant variations in its mass and radius. Assuming the lower mass by L26, the planet loses more than half of its primordial atmosphere and its radius contracts, depending on the model, up to around $4.5\,R_\oplus$, evolving into a mature sub-Neptune. Alternatively, adopting the higher mass from SM21, the deeper gravitational potential ensures that \planet\ remains largely unaffected by photoevaporation.\\
        \indent{Our} findings highlight that without continuous activity monitoring and full transit coverage for all epochs, establishing a robust baseline to disentangle stellar contamination from the true planetary signal remains a formidable challenge. Ultimately, for a star as variable as \starr, co-adding multiple nights with uneven coverage remains problematic. A complete transit with simultaneous multi-wavelength monitoring represents the only definitive path to fully resolve this complex puzzle, requiring dedicated proxies for the stellar wind and XUV emission. In this context, future UV-to-near-infrared space concepts such as the Pollux instrument \citep{pollux} for the Habitable Worlds Observatory (HWO, \citealt{Dressing2026}) will enable a simultaneous evaluation of stellar high-energy behaviour and the planetary \Hei\ response, ultimately unveiling the true nature of signals detected around such active host stars.

        \par

        \begin{acknowledgements}
                We thank the referee for helpful feedback that improved our work. This work is based on observations made with the Italian Telescopio Nazionale Galileo (TNG) operated on the island of La Palma by the Fundaci\'on Galileo Galilei of the INAF at the Spanish Observatorio Roque de los Muchachos of the IAC in the frame of the program Global Architecture of the Planetary Systems (GAPS).
                We thank A.~Vidotto, and A.~Presa for the interesting discussion on star-planet magnetic interactions in photoevaporating exoplanets.
                Based on observations made with the Nordic Optical Telescope, owned in collaboration by the University of Turku and Aarhus University, and operated jointly by Aarhus University, the University of Turku and the University of Oslo, representing Denmark, Finland and Norway, the University of Iceland and Stockholm University at the Observatorio del Roque de los Muchachos, La Palma, Spain, of the Instituto de Astrofisica de Canarias. The NOT data were obtained under program ID 62-009; P.I.: Luigi Mancini. The ALFOSC instrument is provided by the Instituto de Astrofisica de Andalucia (IAA) under a joint agreement with the University of Copenhagen. 
                We acknowledge the Italian center for Astronomical Archives (IA2, https://www.ia2.inaf.it), part of the Italian National Institute for Astrophysics (INAF), for providing technical assistance, services and supporting activities of the GAPS collaboration.
                We acknowledge financial support from the Agencia Estatal de Investigaci\'on of the Ministerio de Ciencia e Innovaci\'on MCIN/AEI/10.13039/501100011033 and the ERDF “A way of making Europe” through projects PID2021-125627OB-C32, CEX2021-001131-S,  PID2022-137241NB-C42, PID2022141216NB-I00 and PID2024-158486OB-C32. This work is also supported by the ERC Grant (ERC Advanced Grant SPEAR, GA 101200674),  funded by the European Union. Views and opinions expressed are however those of the author(s) only and do not necessarily reflect those of the European Union or the  European Research Council Executive Agency (ERCEA). Neither the European Union nor the granting authority can be held responsible for them.
                JS acknowledges support from STFC under grant number ST/Y002563/1.
                FB acknowledges support from INAF GO Large Grant 2024 “Strenghtening pylons for BRIDGES”. The authors acknowledge the support from ASI-INAF agreement 2021-5-HH.0. CDM acknowledges the support of the Next Generation EU funds within the National Recovery and Resilience Plan (PNRR), Mission 4 – Education and Research, Component 2 – From Research to Business (M4C2), Investment Line 3.1 – Strengthening and creation of Research Infrastructures, Project IR0000034 – “STILES – Strengthening the Italian Leadership in ELT and SKA”
        \end{acknowledgements}
        
        \bibliographystyle{aa}
        \bibliography{ref_he}   
        
        \onecolumn
        \begin{appendix}
                \section{Additional figures and tables} \label{additional_tables}
                This section provides additional figures and tables that complement the analysis presented in the main text.\\
                \begin{table*}[ht!]
                        \centering
                        \footnotesize
                        \caption{Comparison of stellar parameters of V1298 Tau and planetary parameters of \planet\ from the literature. \label{tab:parcomp}}
                        \small 
                        \resizebox{\linewidth}{!}{
                                \begin{tabular}{lcccccccc}
                                        \hline
                                        \hline
                                        \noalign{\smallskip}
                                        Reference     &  $M_\star$ & $R_\star$ & $v\sin i_\star$ & $P$ & $R_{\rm p}/R_\star$ & $a/R_\star$ & $b$ & $i$ \\
                                        \noalign{\smallskip}
                                        &  ($M_\odot$) & ($R_\odot$) & (km s$^{-1}$) & (d) &  &  & & (deg) \\
                                        \noalign{\smallskip}
                                        \hline
                                        \noalign{\smallskip}
                                        \noalign{\smallskip}
                                        D19a & $1.10\pm0.05$ & $1.305\pm0.070$ & $23\pm2$ & $24.13889^{+0.00043}_{-0.00044}$ & $0.07111^{+0.00117}_{-0.00061}$ & $28.7^{+1.5}_{-2.3}$ & $0.23^{+0.16}_{-0.15}$ & $89.51^{+0.33}_{-0.38}$ \\
                                        \noalign{\smallskip}
                                        D19a$^{(a)}$ & $1.099^{+0.049}_{-0.049}$ & $1.314^{+0.052}_{-0.064}$ & $\dots$ & $24.13861^{+0.00102}_{-0.00090}$ & $0.0713^{+0.0021}_{-0.0020}$ & $27.6^{+1.5}_{-1.0}$ & $0.29^{+0.16}_{-0.19}$ & $89.35^{+0.41}_{-0.33}$ \\
                                        \noalign{\smallskip}
                                        D19b & $1.101^{+0.049}_{-0.051}$ & $1.345^{+0.056}_{-0.051}$ & $\dots$ & $24.1396\pm0.0018$ & $0.0700\pm0.0023$ & $27.0\pm1.1$ & $0.46^{+0.13}_{-0.24}$ & $89.00^{+0.46}_{-0.24}$  \\
                                        \noalign{\smallskip}
                                        G21 & $1.145\pm0.075$ & $1.317\pm0.027$ & $23\pm2$$^{(b)}$ & $\dots$ & $\dots$ & $27.0\pm1.1$$^{(b)}$ & $\dots$ & $\dots$ \\
                                        \noalign{\smallskip}
                                        V21 & $1.104^{+0.046}_{-0.045}$ & $1.333^{+0.030}_{-0.024}$ & $\dots$ & $\dots$$^{(c)}$ & ${0.0128^{+0.0137}_{-0.0089}}$$^{(d)}$ & $\dots$ & $0.439^{+0.041}_{-0.016}$ & $\dots$ \\
                                        \noalign{\smallskip}
                                        F21$^{(e)}$ & $1.101^{+0.049}_{-0.051}$$^{(f)}$ & $1.345^{+0.056}_{-0.051}$$^{(f)}$ & $23.65^{+0.87}_{-1.17}$ & $\dots$ & $\dots$ & $\dots$ & $\dots$ & $\dots$ \\
                                        \noalign{\smallskip}
                                        F22 & $1.095^{+0.049}_{-0.047}$ & $1.33^{+0.04}_{-0.03}$ & $\dots$ & $24.1315^{+0.0033}_{-0.0034}$ & $0.0636^
                                        \pm0.0018$ & $\dots$ & $0.45^{+0.05}_{-0.04}$ & $\dots$ \\
                                        \noalign{\smallskip}
                                        J22 & $1.135\pm0.047$ & $1.407\pm0.018$ & $24.77\pm0.19$ & $24.141341\pm0.000023$ & $0.071998\pm0.00062$ & $26.06\pm0.46$ & $0.564\pm0.026$ & $88.759\pm0.077$  \\
                                        \noalign{\smallskip}
                                        SM21 & $1.17\pm0.06$ & $1.28\pm0.07$ & $23.8\pm0.5$ & $24.1399\pm0.0015$ & $0.0698\pm0.0024$ & $\dots$ & $\dots$ & $$>88.7$$ \\
                                        \noalign{\smallskip}
                                        F23 & $1.26\pm0.06$ & $1.43\pm0.03$ & $24.9\pm0.5$ & $24.1315$$^{(g)}$ & $\dots$ & $\dots$ & $\dots$ & $\dots$ \\
                                        \noalign{\smallskip}
                                        S23 & $1.157^{+0.057}_{-0.058}$ & $1.355^{+0.032}_{-0.030}$ & $\dots$ & $24.140410\pm0.000022$ & $0.0673\pm0.0017$ & $27.25^{+0.40}_{-0.49}$ & $0.451^{+0.0.033}_{-0.0.030}$ & $89.052^{+0.072}_{-0.087}$ \\
                                        \noalign{\smallskip}
                                        L26$^{(h)}$ & $1.10\pm0.05$$^{(i)}$ & $1.32\pm0.05$$^{(i)}$ & $\dots$ & $24.140006\pm0.000017$ & $\dots$ & $\dots$ & $\dots$ & $\dots$ \\
                                        \noalign{\smallskip}
                                        L26$^{(j)}$ & $\dots$ & $\dots$ & $\dots$ & $24.140420\pm0.000012$ & $0.0661^{+0.0019}_{-0.0017}$ & $\dots$ & $0.26\pm0.17$ & $\dots$  \\
                                        \noalign{\smallskip}
                                        L26$^{(k)}$ & $\dots$ & $\dots$ & $\dots$ & $24.13934\pm0.00030$ & $\dots$ & $\dots$ & $\dots$ & $\dots$ \\
                                        \noalign{\smallskip}
                                        \noalign{\smallskip}
                                        \hline
                                \end{tabular}
                        }
                        \tablefoot{
                                $^{(a)}$ Recommended values based on simultaneous variability and transit fits;
                                $^{(b)}$ Adopted from D19a and D19b;
                                $^{(c)}$ No period value reported. V21 noted that they fixed the mean
                                ephemerides to the best-fit solution from Livingston et al. (2021, in preparation) and that paper was later published as L26;
                                $^{(d)}$ Typo in Table 2 of V21, in which the radius ratio value of V1298 Tau c is used. The likely correct value for \planet\ is $R_{\rm p}/R_\star = 0.0642^{+0.0048}_{-0.0047}$, which is erroneously reported for V1298 Tau d;
                                $^{(e)}$ The F21 work focusses on V1298 Tau c, but it is relevant for the independent estimate of $v\sin i_\star$ for the primary;
                                $^{(f)}$ Adopted from D19b;
                                $^{(g)}$ Adopted from F22;
                                $^{(h)}$ Based on analytic TTV modeling reported in Table 1 and Supplementary Table 3 of L26;
                                $^{(i)}$ Adopted from D19a;
                                $^{(j)}$ Based on joint modeling of the K2 and TESS photometry reported in Extended Data Table 1 of L26;
                                $^{(k)}$ Based on results of N-body TTV modeling reported in Extended Data Table 2 of L26.
                        }
                \end{table*}
                
                \begin{table}[ht!]
                        \centering
                        \renewcommand{\arraystretch}{1.5}
                        \caption{Results from \textsc{ATES} simulations showing the mass loss rate, transit depth and EW as a function of the \text{XUV} flux relative to the quiet star's flux ($F_0$).}
                        \label{tab_ates}
                        \begin{tabular}{c ccc ccc}
                                \hline \hline
                                &
                                \multicolumn{3}{c}{SM21: $M_{\rm p} = 0.63 M_{\rm Jup}$}&
                                \multicolumn{3}{c}{L26: $M_{\rm p} = 0.042 M_{\rm Jup}$}\\
                                \hline
                                $F_{\text{XUV}}/F_0$                   &
                                $\dot{M} (\text{g~s}^{-1})$ &
                                c ($\%$)                               &
                                EW(\Hei) ($\text{m\AA}$)    &
                                $\dot{M} (\text{g~s}^{-1})$ &
                                c ($\%$)                               &
                                EW(\Hei) ($\text{m\AA}$)   \\
                                \hline
                                $0.5$  & $8.13 \times 10^{9}$  & $5.90$ & $29.43$ & $2.24 \times 10^{12}$ & $15.60$ & $154.9$ \\
                                $1.0$  & $1.62 \times 10^{10}$ & $4.78$ & $25.90$ & $3.98 \times 10^{12}$ & $20.58$ & $231.6$ \\
                                $2.0$  & $3.16 \times 10^{10}$ & $3.72$ & $21.98$ & $6.92 \times 10^{12}$ & $26.00$ & $328.3$ \\
                                \hline
                        \end{tabular}
                \end{table}
                \twocolumn

                \begin{figure}
                        \centering
                        \includegraphics[height=4cm]{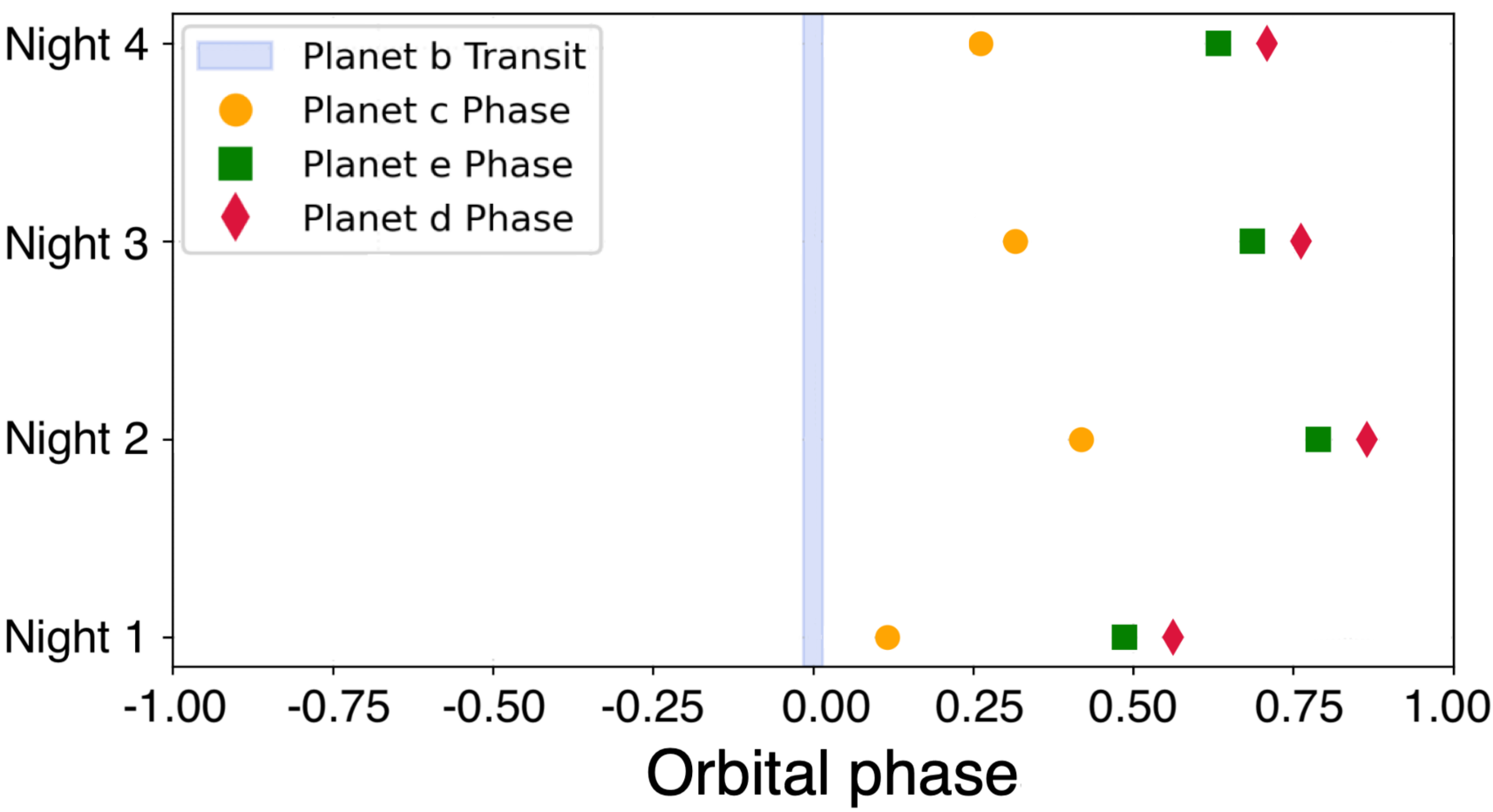}
                        \caption{
                                Position of the V1298\,Tau's planets during each observed night. The closest planet is planet~c, which transited 2 days before \nuno.}
                        \label{orbite}
                \end{figure}
                \begin{figure}
                        \centering
                        \includegraphics[height=3.5cm]{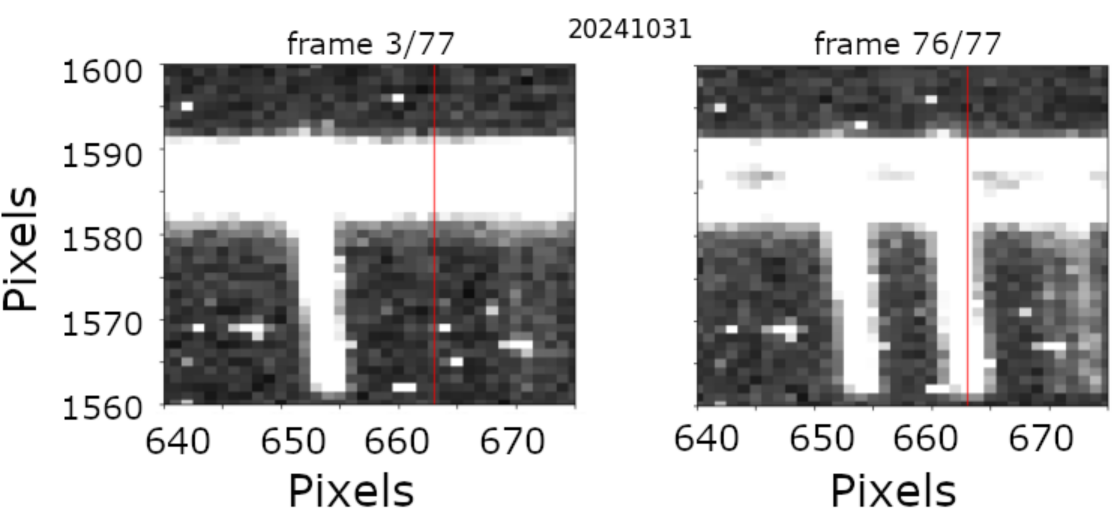}
                        \caption{
                                Raw spectra at the beginning (\textit{left}) and at the end (\textit{right}) of \nquattro\ around the helium triplet. 
                                Both images were taken with the same nodding position. 
                                In the first frame, the strong telluric OH doublet (10834.241~\AA\ and 10834.338~\AA, vacuum wavelengths) is clearly visible, while in the second frame, a second emission line appears.  The vertical red line indicates the position of this emission feature, allowing comparison between the two frames. 
                                This line is airglow from metastable helium at 1.083~$\mu$m.}
                        \label{emission_line}
                        
                \end{figure}

                \begin{figure}
                        \centering
                        \includegraphics[width=0.89\linewidth]{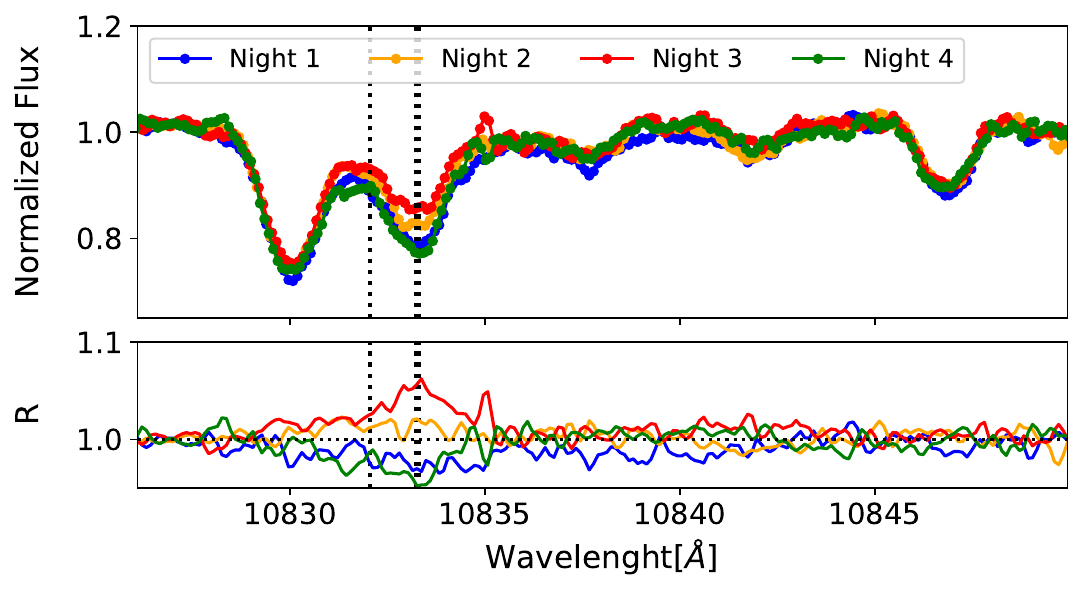}
                        \includegraphics[width=0.89\linewidth]{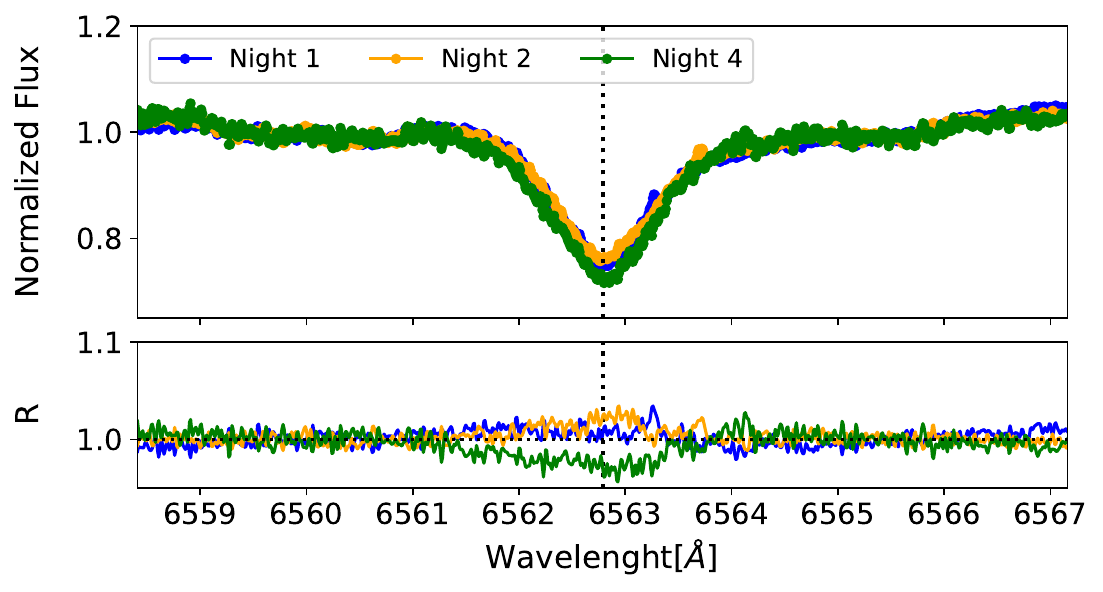}
                        \caption{Time-averaged out-of-transit spectra in the stellar rest frame for the \Hei\ triplet (\textit{top}) and the \Ha\ line (\textit{bottom}). Vertical black lines mark the rest-frame positions of the respective features. For each line, the ratio $R$ relative to the global master-out spectrum is also shown. Significant night-to-night variability is clearly visible.}
                        \label{so_out}
                \end{figure}

                \begin{figure} [h!]
                        \centering
                        \includegraphics[width=0.9\linewidth]{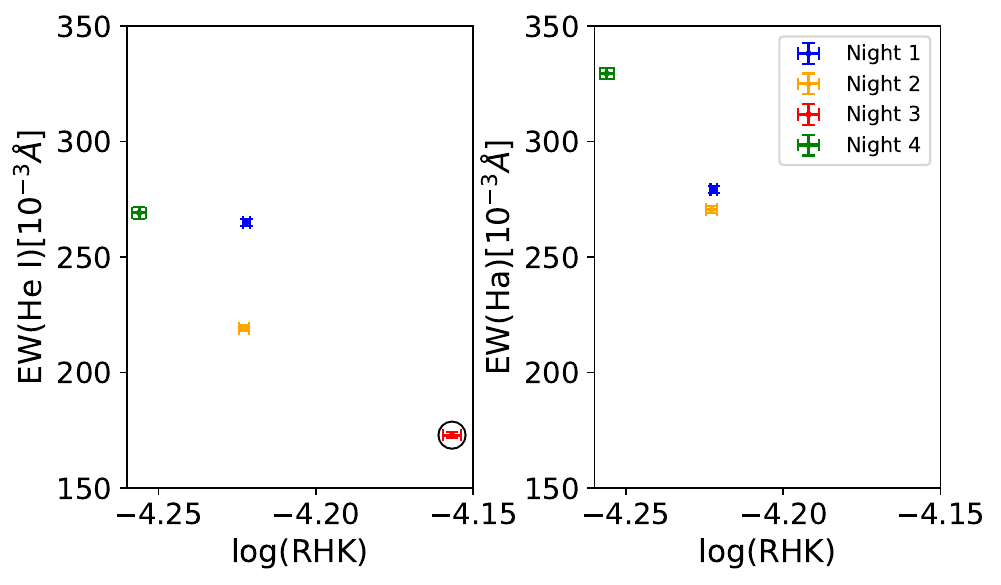}
                        \includegraphics[width=0.45\linewidth]{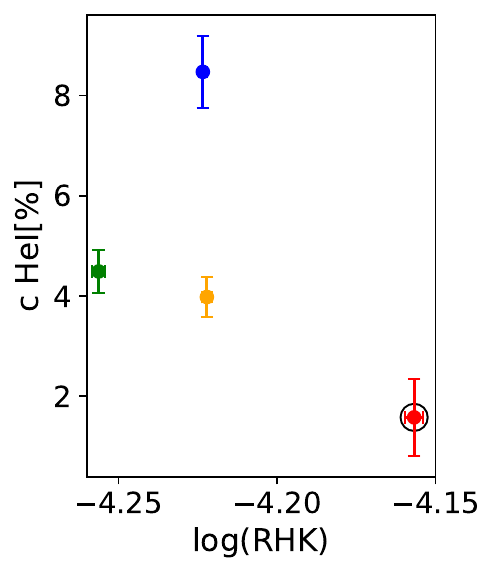}
                        \caption{\textit{Top panels}: EWs of the master out-of-transit spectra as a function of the average out-of-transit \logrhk\ for the \Hei\ (left) and \Ha\ (right) lines. \textit{Bottom panel}: \Hei\ absorption contrast ($\%$) as a function of \logrhk. For \ntre\ (highlighted by a black circle), the \logrhk\ value represents the average of the in-transit measurements, as no out-of-transit data were acquired for this indicator during that epoch.}
                        \label{logrhkew}
                \end{figure}
                \begin{figure}[h!]
                        \centering
                        \includegraphics[width=0.9\linewidth]{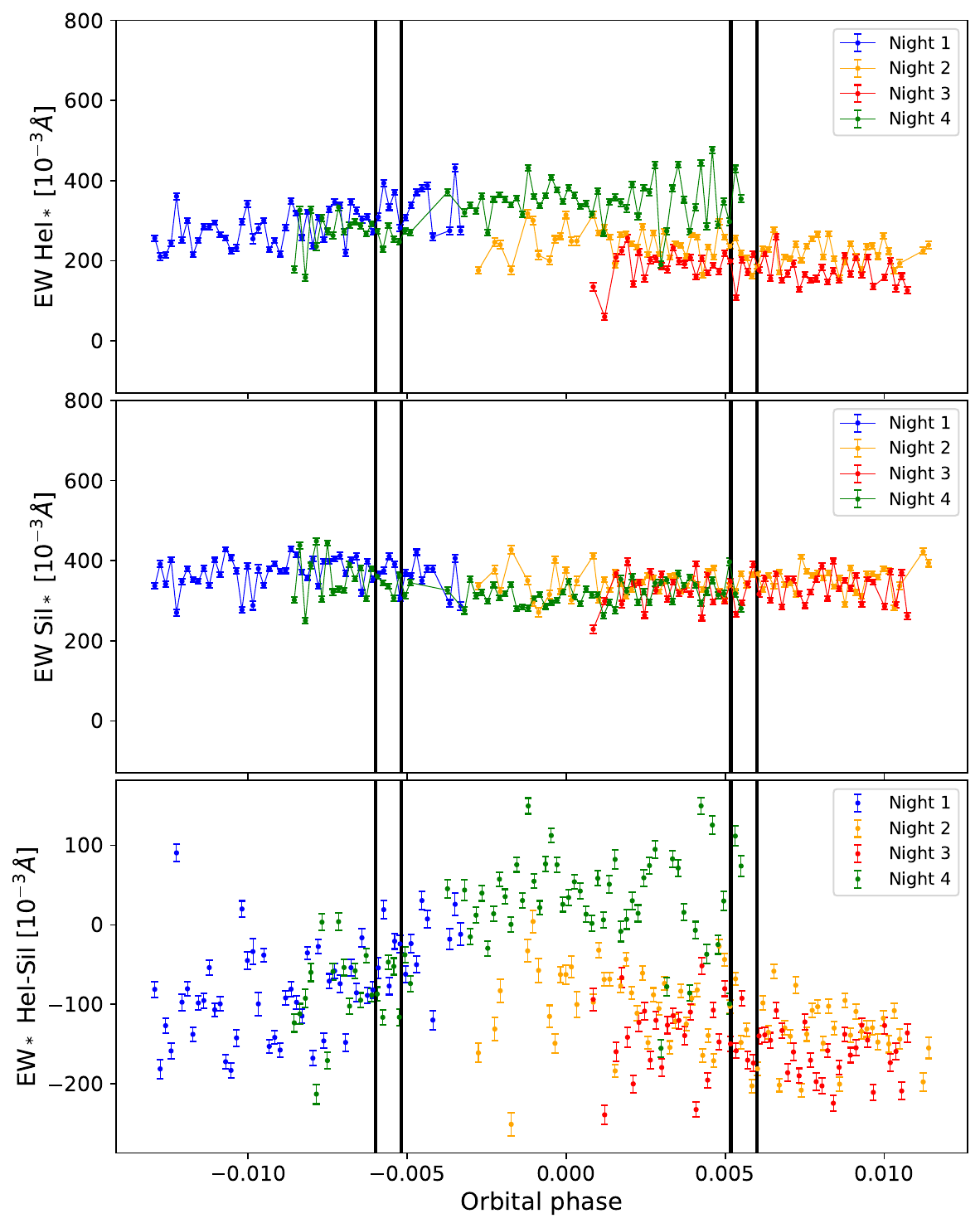}      \caption{EW of the stellar \Hei\ (\textit{top}) and \Sii\ (\textit{middle}), and the difference between the \Hei\ and \Sii\ EWs (\textit{bottom}).}
                        \label{EW_star}
                \end{figure}
                \twocolumn

                \begin{figure}
                        \centering
                        \includegraphics[width=\linewidth]{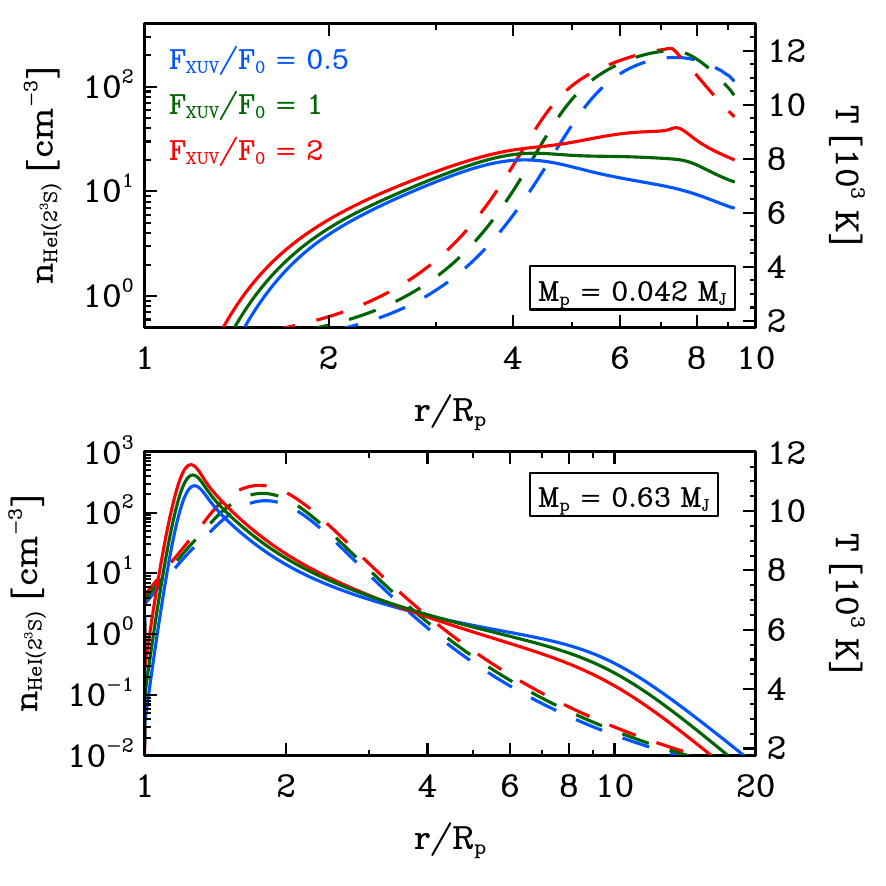}
                        \caption{Helium triplet number densities (thick line, left axis) and temperature (dashed, right axis) as a function of radial distance from the planet's center obtained from the ATES simulations, assuming L26's (\textit{top}) and SM21's (\textit{bottom}) value for the mass of the planet. Different colours indicate different irradiation levels (cf. Table~\ref{tab_ates}).}
                        \label{fig:ATESHeTR}
                \end{figure}

                \section{NOT light curve}
                \label{not}
                A transit of V1298\,Tau\,b was photometrically monitored with the NOT telescope on December 7, 2020, simultaneously with GIARPS observations on night 2. We used the Alhambra Faint Object Spectrograph and Camera (ALFOSC) in its imaging mode and selected the engineered diffuser $\#1$ \citep{vonEssen2019}. The observations were performed through a $I$-band ($\# 12$) filter. The $6.5^{\prime}\times6.5^{\prime}$ FOV was suitably chosen to include three good reference stars, namely HD\,284152, HD\,284153, and HD\,284154. The observations started at 19:35 UT and ended at 04:39 UT. The night started with a few clouds, which cleared by 23:00\,UT. The airmass varied from 2.28 to 1.01 and then to 1.98. The moon illuminated fraction at the midpoint of exposure was $50\%$. The exposure time was 35\,s at the beginning of the observations and was gradually reduced to 6\,s. A total of 918 images were collected.

                The data were analysed using the {\sc defot} pipeline
                \citep{southworth2009,southworth2014}. Following a standard approach,
                which has been used in many previous works (e.g. \citealt{southworth2012,southworth2015}), the raw science images were initially calibrated using master-bias and master-flat frames. These calibration frames were created by median-combining individual bias and sky flat-field images captured just after sunset. Subsequently, we performed aperture photometry by selecting a set of five comparison stars and cross-correlating each image in the time series with a reference frame. We tried different aperture sizes and retained those (22\,px, 45\,px, 65\,px) that gave the light curve with the smallest scatter. Finally, differential-magnitude light curves were extracted relative to each comparison star. Instrumental magnitudes were converted into differential-magnitude light curves, which were normalized to zero out of transit. This normalization was achieved by fitting second-order polynomials to the out-of-transit data. The differential magnitudes were calculated relative to a weighted ensemble of five comparison stars that we already mentioned. To minimize the scatter in the out-of-transit data, we simultaneously optimized both the comparison star weights and the polynomial coefficients. Timestamps were converted from \(\text{JD}_{\text{UTC}}\) to \(\text{BJD}_{\text{TDB}}\) following the recommendations of \citet{eastman2010}.
                
\begin{figure}
                        \centering
                        \includegraphics[width=0.9\linewidth]{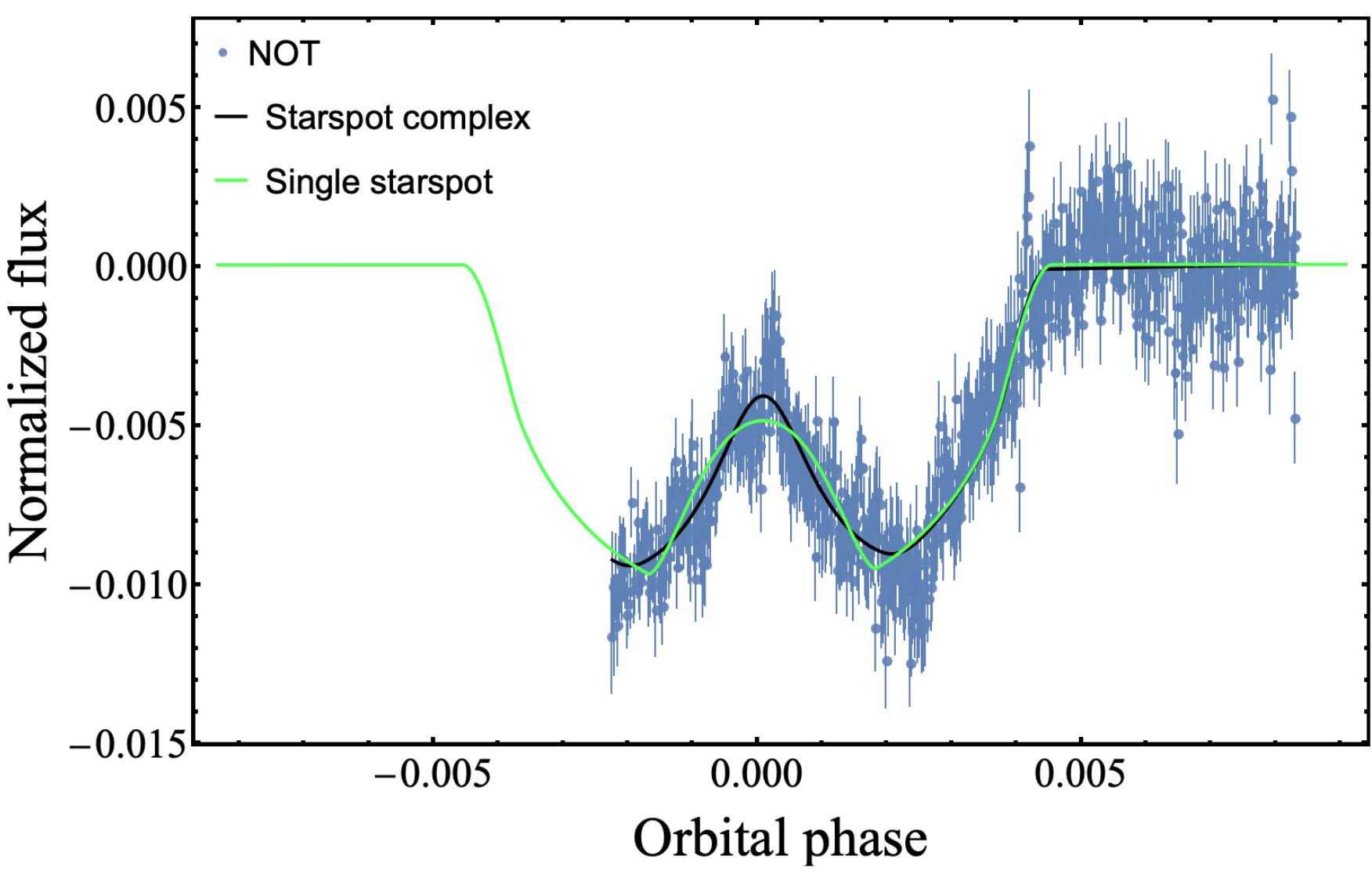}
                        \includegraphics[width=0.9\linewidth]{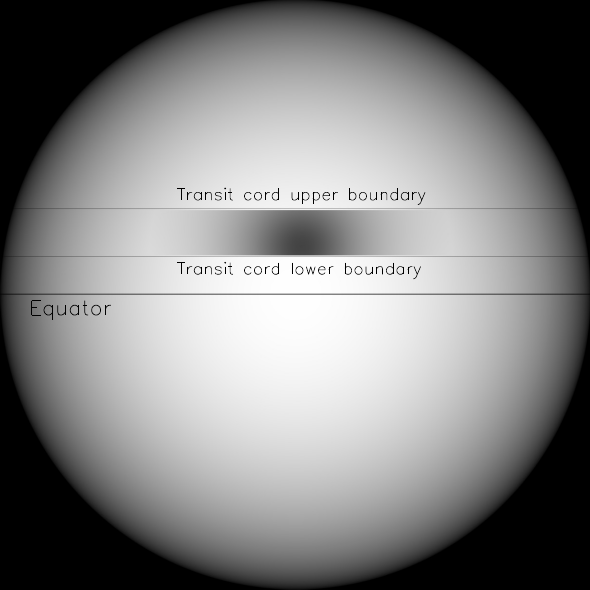}
                        \caption{{\it Top}:
                                Light curve of the V1298\,Tau\,b transit, which we observed with AFOSC@NOT, compared with the best \texttt{prism+gemc} fits. {\it Bottom}: Representation of the stellar disc, starspot-complex position, and transit chord for the dataset containing a starspot anomaly.
                        }
                        \label{NOT_lc}
                \end{figure}

                Due to adverse weather conditions in the early part of the night, the first data points of the light curve resulted in being very scattered. Therefore, we were forced to exclude the first 222 points from the analysis and considered only those starting from 22:32\,UT, thereby losing all the ingress data points. The final light curve, composed of 696 data points, presents a clear bump that peaks close to the transit midpoint and lasts about half the entire transit. We interpreted this anomaly as the consequence of a starspot complex on the stellar photosphere that was occulted by the planet during its transit; top panel of Fig.~\ref{NOT_lc}.
                
                Following the approach of previous works in which similar cases have come across (e.g. \citealt{mohlerfischer2013,mancini2014,mancini2015,mancini2017,biagiotti2024}), we used the \texttt{prism} and \texttt{gemc} codes\footnote{\url{https://github.com/JTregloanReed/PRISM_GEMC}\,.} \citep{tregloanreed2013,tregloanreed2015}, which allowed us to model the light
                curve by fitting both the transit and the starspot-crossing event simultaneously, thanks to a Bayesian approach and the use of genetic algorithms \citep{TerBraak2006}. Utilizing a quadratic limb-darkening (LD) law and assuming a circular orbit, \texttt{prism+gemc} models light curves by fitting the following parameters: the sum of the fractional stellar and planetary radii ($r_{\rm p}+r_{\star}$, $r_{\rm p}$ and $r_{\star}$ are defined as the planetary radius  $R_{\rm p}$ and the stellar radius $R_{\star}$ scaled by the semi-major axis $a$, respectively), their ratio $r_{\rm p}/r_{\star}$, the orbital inclination $i$ and the LD coefficients. It also fits the mid-transit time, the longitude and latitude of the starspot centre ($\phi,\,\theta$), its radius $r_{\rm s}$, and its contrast $\rho$, the latter defined as the surface brightness ratio between the starspot and the surrounding photosphere.
                
                We started the modelling of the light curve by fixing the orbital period at $P_{\rm orb} = 24.141441$\,days and the mid-transit time at $T_0=2\,459\,191.4932383439$ (private communication), and assuming the starspot as a spherical cap. By imposing this geometry, a large ($25^{\circ}\pm4^{\circ}$), high-latitude starspot at $48^{\circ}\pm5^{\circ}$ is needed to reproduce such a long-lasting bump. Although it may be a long active region located along the transit chord, such a large starspot is very suspicious because it would produce an apparent blueshift as it moves across the half of the stellar disk that is moving away from the observer. In reality, \texttt{prism+gemc} scans the stellar surface within the transit chord, which, when using circular starspots, causes (in some cases) large starspots that graze the transit chord. Therefore, we tried a different approach and attempted to model an elongated series of small starspots instead of a single starspot. We imposed a starspot on the pixels within the transit cord and apply a contrast gradient to the starspot. The gradient follows a cosine function where the starspot centre has a contrast given by the best-fitting parameter and the starspot edge has a contrast of 1.0. The equation we used is
                \begin{equation}
                        |\vec\nabla{\rho}| =1 - \cos{\xi} \, (1 - \rho)\,,
                \end{equation}
                where $\xi$ is the angle between the spot centre and the pixel being evaluated and ranges between 0 (starspot centre) and 90 degrees (starspot edge). The best-fitting parameters indicate a starspot complex within the transit cord, whose sizes are approximately 29 degrees in length and 9 degrees in height, bottom panel of Fig.~\ref{NOT_lc}. The parameters of both fits (single starpsot and starspot complex) are reported in Table~\ref{tab:not_analysis}.
                \begin{table}[t]
                        \centering
                        \caption{Photometric properties of the V1298\,Tau system and starspot parameters derived by fitting the NOT light curve with \texttt{prism+gemc}. }
                        \label{tab:not_analysis}
                        \begin{tabular}{lcc}
                                \hline
                                \hline  \\[-8pt]
                                Parameter &  Single   &  Starspot \\
                                &  starspot &  complex  \\ [2 pt]
                                \hline  \\[-6pt] 
                                $R_{\rm p}/R_{\star}$  & $0.0823 \pm 0.0020$ & $0.0807 \pm 0.0080$ \\
                                $a/R_{\star}$  & $33.8 \pm 2.5$ & $33.4 \pm 7.0$ \\
                                $i$\,(deg)  & $89.23 \pm 0.18$ & $89.58 \pm 0.39$ \\
                                $u_{\rm A}$  & $0.93 \pm 0.17$ & $0.990 \pm 0.047$ \\
                                $v_{\rm A}$  & $0.07 \pm 0.16$ & $0.010 \pm 0.045$ \\
                                $\theta$\,(deg)  & $1.11 \pm 0.45$ & $1.13 \pm 0.36$ \\
                                $\phi$\,(deg)  & $48.2  \pm 4.8$ & $9.3 \pm 9.9$ \\
                                $r_{\rm spot}$\,(deg)$^{(a)}$  & $24.6 \pm 3.8$ & 
                                \begin{tabular}{c} 
                                        $29\pm 10$ \\
                                        $ 9.24 \pm 0.92$ \\ 
                                \end{tabular}
                                \\
                                $\rho_{\rm spot}$  & $0.338 \pm 0.050$ & $0.270 \pm 0.070$ \\
                                $T_{\rm spot}$\,(K)  & $3890 \pm 136$ & $3711 \pm 204$ \\
                                \hline
                        \end{tabular}
                        \tablefoot{$^{(a)}$In the starspot complex scenario, two values are reported because the complex is modeled with a rectangular geometry (requiring two angular dimensions) rather than a circular shape.}
                \end{table}

                \section{\ion{Na}{I} analysis} \label{sodium_app}
                For the \ion{Na}{I} analysis, we employed the \texttt{SLOPpy} (Spectral Lines Of Planets with python) pipeline \citep{Sicilia2022}, a standard user-friendly and reliable tool optimised for spectral reduction and the extraction of planetary transmission spectra in the VIS obtained from high-resolution
                observations. Although the main body of this work investigates a total of four observational nights, the sodium line analysis was exclusively performed on the first two nights, which are the only ones featuring high-resolution data from the HARPS-N spectrograph. During the data reduction and selection process, one out-of-transit spectrum from night 1 was discarded due to an extremely low signal-to-noise ratio ($\text{S/N} \approx 2.6$) to prevent the introduction of severe systematics or spurious noise into the final transmission spectrum.
                
                The spectra were corrected for stellar spectroscopic effects that typically contaminate transit observations, namely the Rossiter-McLaughlin (RM) effect and center-to-limb variations (CLV). These corrections were implemented directly through \texttt{SLOPpy}, utilizing synthetic spectral models generated via \texttt{PySME} \citep{PySME}. The stellar atmosphere models were computed using the line list from the third version of the Vienna Atomic Line Database (VALD3\footnote{\url{https://vald.astro.uu.se}}) combined with MARCS stellar atmosphere models, adopting non-LTE grids.
                
\begin{figure}
                        \centering
                        \includegraphics[width=0.9\linewidth]{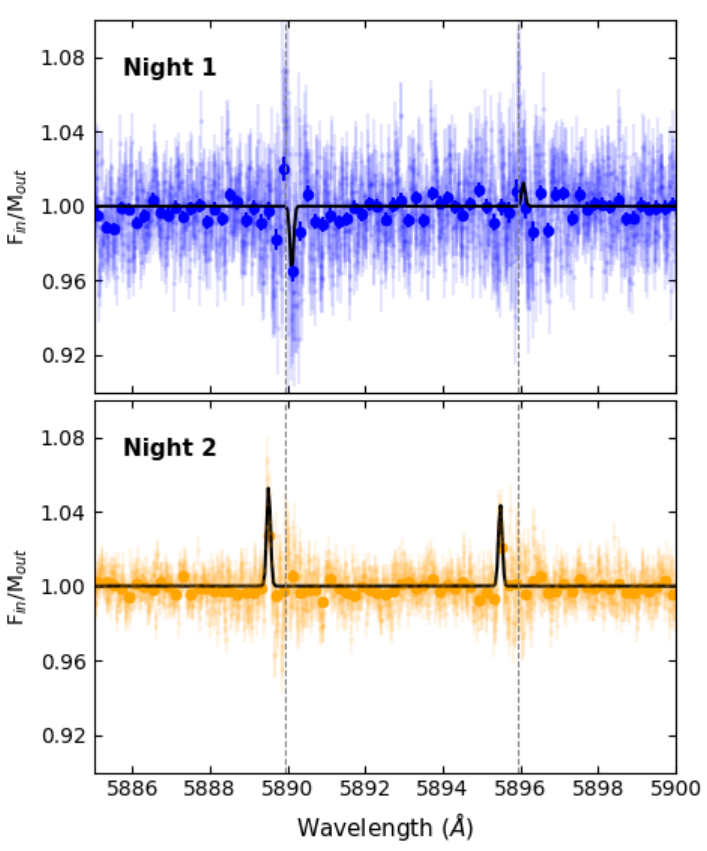}
                        \caption{Transmission spectra of \planet\ centred around the \ion{Na}{I} doublet in the planetary rest frame for night 1 (\textit{top}) and night 2 (\textit{bottom}). The small, semi-transparent points in the background represent the full-resolution observational data, whereas the distinct circles with their corresponding error bars indicate the data binned 20 times.
                                The solid black line is the MCMC Gaussian fit performed by \texttt{SLOPpy}, while the vertical dashed gray lines indicates the rest frame transition wavelengths of the \ion{Na}{I} D lines.}
                        \label{fig:Na_transmission_spectra}
                \end{figure}

                The final transmission spectrum, corrected for stellar systematics, was modeled using a Markov Chain Monte Carlo MCMC sampling procedure within \texttt{SLOPpy}, applying a prior on the theoretical planetary orbital radial velocity semi-amplitude ($K_{\rm p}$). The model assumes a double Gaussian profile for the \ion{Na}{i} atmospheric absorption or emission lines and a flat spectrum otherwise. The free parameters allowed to vary during the fitting process include $K_{\rm p}$, which is required to model the atmospheric lines in the planetary rest frame; the line contrasts ($c_{\text{D1}}$ and $c_{\text{D2}}$); the FWHM of the double Gaussian profile; and the residual radial velocity of the atmospheric wind ($v_{\text{wind}}$), defined as the shift relative to the transition in the planetary rest frame. We assumed a shared FWHM and $v_{\text{wind}}$ for both D lines. To account for any unmodeled instrumental or environmental systematics, such as variations in seeing, a free jitter parameter ($\text{jitter}_0$) was also d in the fitting procedure. The median values of the resulting posterior distributions were adopted as the best-fit parameters, with error bars corresponding to the $1\sigma$ statistical uncertainties at the 15th and 84th percentiles. The final transmission spectra are shown in Fig. \ref{fig:Na_transmission_spectra}. 
                
                \begin{table}
                        \centering
                        \caption{Best-fit MCMC parameters obtained with \texttt{SLOPpy} for the \ion{Na}{i} doublet across the two observational nights.}
                        \label{tab:na_results}
                        \begin{tabular}{lcc}
                                \hline
                                \hline
                                Parameter & Night 1 & Night 2  \\
                                \hline
                                $c_{\text{D1}} (\%)$ & $+1.3^{+3.8}_{-2.2}$ & $+4.4^{+0.5}_{-0.7}$ \\
                                $c_{\text{D2}}$ (\%) & $-3.5^{+7.3}_{-3.2}$ & $+5.3^{+0.6}_{-0.6}$ \\
                                $\text{FWHM}_{\text{shared}}$ ($\text{km}\,\text{s}^{-1}$) & $5.6^{+3.5}_{-5.0}$ & $5.8^{+0.5}_{-0.6}$ \\
                                $v_{\text{wind}}$ ($\text{km}\,\text{s}^{-1}$) & $+7.9^{+3.5}_{-6.9}$ & $-22.48^{+0.35}_{-0.24}$ \\
                                $K_{\rm p}$ ($\text{km}\,\text{s}^{-1}$) & $76.7^{+10.8}_{-9.3}$ & $52.6^{+10.0}_{-9.3}$ \\
                                $\text{jitter}$ & $0.0200 \pm 0.0008$ & $0.0193 \pm 0.0004$ \\
                                \hline
                        \end{tabular}
                \end{table}
                
                The MCMC fit results for the two nights reveal a significant discrepancy in signal quality and parameter constraints, as summarised in Table~\ref{tab:na_results}. For night 1, the retrieved parameters exhibit exceptionally large uncertainties. Although the nominal best-fit profile suggests a disparate behaviour between the two lines—with $\text{D}_2$ in absorption and $\text{D}_1$ in emission—both contrasts are fully compatible with zero within their error bars. Consequently, the data from night 1 are inconclusive and can neither confirm nor rule out an atmospheric signal. Conversely, night 2 yields a much cleaner and tightly constrained signal, characterised by a distinct double emission profile for both components of the doublet. The positive contrast values robustly confirm an emission signal well above the noise floor, accompanied by a strong blue-shift.

                We note that this pronounced discrepancy between the two observational epochs closely follows the variability observed in the H$\alpha$ line. Conversely, the \Hei\ triplet displays a consistent behaviour, with transit absorption being detected on both nights.

                \section{Modelling an absorption leading tail in V1298~Tau~b}
                \label{appendix_wind}
                
                A young planet such as V1298~Tau~b is highly inflated and is probably characterized by a strong hydromagnetic dynamo action because of the intense convective flows in its interior. The timescale for the synchronization of its rotation with the orbital motion, $\tau_{\rm sync}$, can be estimated by means of the tidal model of \citet{Leconteetal10}, adopting $k_{2}\Delta t = 3/(2nQ^{\prime})$, where $k_{2}$ is the planet Love number of an order of 2, $\Delta t$ the tidal time lag, $n$ the orbital mean motion, and $Q^{\prime}$ the modified tidal quality factor of the planet that we assume similar to that of Jupiter, $Q^{\prime} \sim 10^{5}$. This gives a timescale of $\tau_{\rm synch} \sim 0.3$~Myr, adopting a gyration radius similar to Jupiter's. This is much shorter than the age of the star, which is estimated to be between 10 and 30 Myr; therefore, the planet's rotation period can be assumed to be equal to its orbital period. Nonetheless, the planetary dynamo can be remarkably strong owing to the large internal thermal energy of the planet responsible for its highly inflated radius \citep{ReinersChristensen10,Christensen10,YadavThorngren17}. As a consequence, we believe that a planetary dipole field with an intensity of $B_{\rm p0} = 25$~G at the poles can be a reasonable assumption. On the other hand, the stellar dipolar magnetic field at the stellar poles can be assumed to be $B_{\rm s0} = 200$~G, according to the Zeeman Doppler Imaging maps of V1298~Tau \citep[cf.][]{Finocietyetal23}. 
                
                The average radius of the planetary magnetosphere can be estimated by assuming pressure equilibrium at the boundary of the magnetosphere in front of the stellar wind flow. Specifically, we assume that the magnetic pressure of the planetary field at the boundary of radius $r_{\rm M}$, measured from the centre of the planet, is equal to the sum of the thermal pressure of the stellar wind, $p_{\rm w}$, the magnetic pressure of the stellar wind field, $B_{\rm w}$, and the ram pressure of the stellar wind having a density $\rho_{\rm w}$ and a relative velocity $u_{\rm w}$ at the distance $a$ of the planet
                \begin{equation}
                        \frac{B_{\rm p}^{2}(r_{\rm M})}{2\mu} \simeq p_{\rm w}(a) + \frac{B_{\rm s}(a)}{2\mu} + \frac{1}{2} \rho_{\rm w}(a) u^{2}_{\rm w}(a), 
                        \label{mag_equil}
                \end{equation}
                where $\mu$ is the magnetic permeability of the plasma that can be assumed to be equal to that of the vacuum. Both the stellar and the planetary fields can be assumed to scale as a dipole field with the distance from the centre of the star and of the planet, respectively. This gives $B_{\rm p} (r_{\rm M}) = B_{\rm p0} (r_{\rm M}/R_{\rm p})^{-3}$ and $B_{\rm s}(a) = B_{\rm s0} (a/R_{\rm s})^{-3}$, where $R_{\rm p}$ and $R_{\rm s}$ are the radius of the planet and of the star, respectively. 
                
\begin{figure}
                        \centering
                        \includegraphics[width=\linewidth]{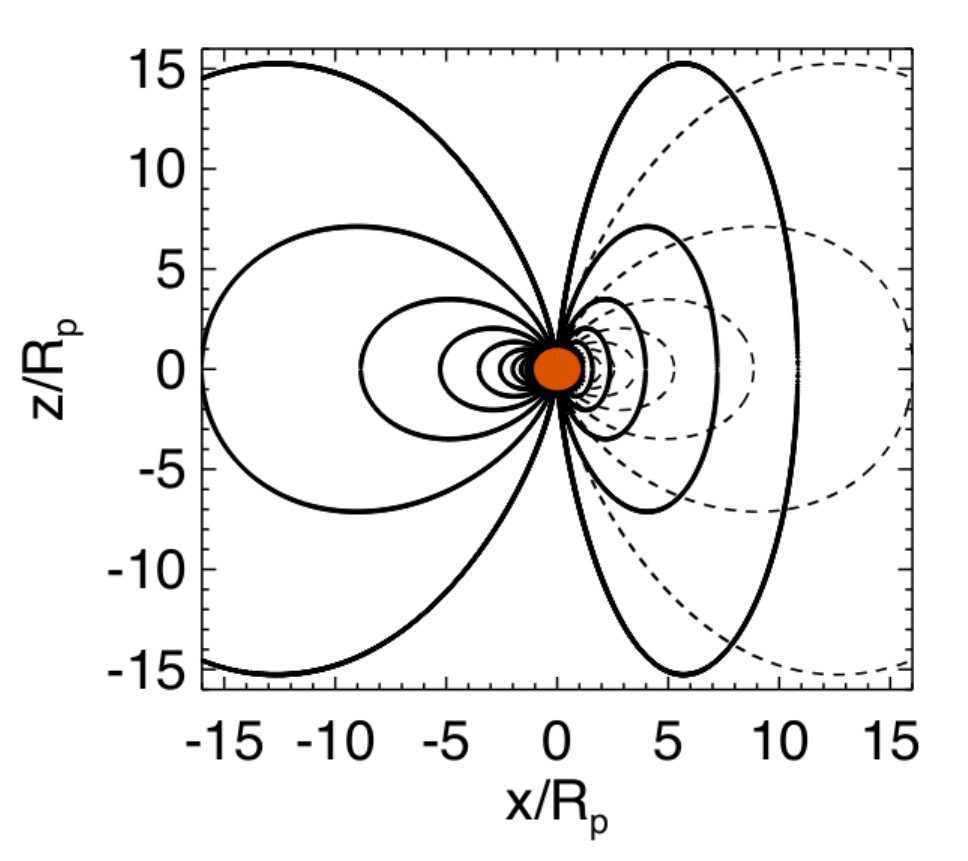}
                        \caption{Schematic section of a planetary magnetosphere with the x-axis pointing towards an  incoming stellar wind streamer and the z-axis perpendicular to the orbital plane of the planet. The planetary dipole axis is assumed to be directed along the z-axis. The magnetospheric field is compressed in the hemisphere facing the streamer because of the wind ram pressure. On the other hand, the planetary field is unperturbed on the opposite hemisphere pointing away from the streamer. The solid lines represent the magnetic field lines in the case of the magnetosphere compressed by the streamer, while the dashed lines  show the unperturbed configuration of the magnetic field, symmetric of that on the opposite unperturbed hemisphere.}
                        \label{planetary_magnetosphere}
                \end{figure}

                Here, we assume that the thermal pressure of the stellar wind is dominated by the pressure of the magnetic field of the wind or by the ram pressure, so that $p_{\rm w}(a) \ll \max[B_{\rm s}^{2}(a)/2\mu, \; \rho_{\rm w}(a) u^{2}_{\rm w}(a)/2]$. In such a case, the planet would need to be relatively close to its host star. The kinetic ram pressure of the wind can be expressed by introducing the Alfven Mach number, $M_{\rm A}$, of the wind at the distance of the planet
                \begin{equation}
                        M_{\rm A} \equiv \frac{u_{\rm w}(a) \sqrt{\mu \rho_{\rm w}(a)}}{B_{\rm s}(a)}.
                \end{equation}
                Substituting this into Eq.~\eqref{mag_equil}, we get 
                \begin{equation}
                        \frac{r_{\rm M}}{R_{\rm p}} \sim (1 + M_{\rm A})^{-1/6} \left(\frac{B_{\rm p0}}{B_{\rm s0}}\right)^{1/3} \left( \frac{a}{R_{\rm s}}\right).  
                        \label{magn_radius}
                \end{equation}
                We can assume that the stellar wind of V1298~Tau is highly inhomogeneous and remarkably time dependent with streamers of widely different velocity and density. 
                Considering a typical situation where the planet is close to the transition between the sub-Alfvenic and the super-Alfvenic regimes of the stellar wind, $M_{\rm A} \sim 1$, and Eq.~\eqref{magn_radius} gives $r_{\rm M}/R_{\rm p} \sim 12.2$. On the other hand, when the planet is reached by a fast and dense streamer of the stellar wind with $M_{\rm A} \sim 10$, the radius of the hemisphere of the magnetosphere impacted by the streamer is compressed to $r_{\rm M}/R_{\rm p} \sim 6.35$, according to Eq.~\eqref{magn_radius}.
                A simple sketch of the compressed planetary magnetosphere is plotted in Fig.~\ref{planetary_magnetosphere} where the star is located at a large distance, $a/R_{\rm p} \sim 400$,  approximately along the positive x-axis. 
                
                The planet V1298~Tau~b is subject to an intense photoevaporation due to the strong XUV radiation coming from its young and active star. The ionized evaporation flow moves along the magnetic field lines of the planetary magnetosphere and fills it with plasma. Considering a typical temperature of $\sim 10^{4}$~K, and the radius and mass of the planet as determined by L26, the pressure scale height is $H_{\rm p }\sim 1.6\; R_{\rm p}$, suggesting that the evaporated plasma fills the magnetosphere up to its outer boundary considered above. Specifically, for that temperature, the density variations between the planet surface and the limit of the magnetosphere are within a factor of about two, while in a planet with a stronger surface gravity remarkably larger variations will occur (Sect.~\ref{plasma_distr}). 
                
                The slow rotation of the planet induced by tidal synchronization implies that the centrifugal force is negligible in comparison to gravity and plasma pressure in the planetary magnetosphere, contrary to the case of Jupiter that is a fast rotator in this context. Quantitatively, even at the outer boundary of the magnetosphere of V1298~Tau~b, $r=r_{\rm M} = 12.2\, R_{\rm p}$, the ratio of the centrifugal to the gravitational acceleration is 0.69 and scales as $r^{3}$ with the distance from the centre of the planet within the magnetosphere. 
                
                The mean density of the magnetospheric plasma in front of the planet, in the hemisphere impacted by the wind streamer, is approximately proportional to the inverse of the volume of the hemisphere itself. Assuming that the cross-section of the hemisphere orthogonal to the wind streamer remains approximately constant, that volume is proportional to $r_{\rm M}$. Therefore, it decreases by a factor of $\sim 2$ when a wind streamer with $M_{\rm A} \sim 10$ impacts onto the magnetosphere of our planet producing an increase in the plasma mean density by a factor of 2. That increase in the density of the magnetospheric plasma in front of the planet can be invoked to interpret the overdensity of the neutral helium as observed in the tail preceding the planet before its transit. The streamer is not moving radially away from the star, rather it forms an angle with the radial direction that depends on the conservation of the angular momentum of the streamer's plasma, its radial velocity, and the orbital velocity of the planet. As a consequence, the compressed magnetosphere can lead the planet along its orbit, thus producing an absorption in the helium triplet that is observable before the ingress of the planet on the stellar disc. 
                
                The present magnetospheric model assumes a pressure balance at the boundary of the planet magnetosphere and a negligible penetration of the stellar wind plasma inside the magnetosphere itself. On the other hand, if the wind plasma penetrates into the magnetosphere, we have to model the magnetospheric boundary as a strong MHD perpendicular shock because of the relatively high Alfven Mach number ($M_{\rm A} \sim 10$). In that case, the density inside the magnetosphere can increase up to a factor of $\sim 4$ \citep[][]{Priest82}, thus further enhancing the absorption in the leading tail over the effect of the magnetospheric compression considered above, while the velocity of the plasma past the shock can lead to an additional Doppler shift and an increase in the width of the absorption of the helium triplet owing to its velocity component in the direction of the observer and the extension of the flow to the hemisphere of the magnetosphere past the planet.
                
                The neutral helium atoms are formed  by recombination of the single-ionized helium and, therefore, their density value is the same as the density of those ions. Nevertheless, once neutral helium atoms form, their dynamics is no longer constrained by the magnetic field and they can freely flow evaporating away from the planet. The helium triplet line blueshifts and widths reflect such an evaporating motion (cf. Fig~\ref{additional}). The blueshift is produced by the component of the evaporation flow towards the observer that is at its maximum during \nuno\ when the evaporation is likely faster because of the higher level of stellar activity. The sonic radius in the case of our very inflated planet is only $\sim 0.3\,R_{\rm p}$ for a temperature of the evaporation flow of $10^{4}$~K implying that the wind flow becomes remarkably supersonic within the boundary of the magnetosphere.
                However, even in this regime, the line broadening that can be explained by the evaporating flow is still of the order of $\pm c_{\rm s}$, where $c_{\rm s}\sim 10$~km~s$^{-1}$ is the sound speed in the evaporating flow. Nevertheless, the line broadening during  \nuno\ is larger by a factor of at least two, suggesting that there is a component of the neutral helium absorption produced by atoms originated by charge exchange between the helium ions of the stellar wind, having a remarkably larger velocity, and the neutral helium atoms at the boundary of the planetary magnetosphere. This combination of absorption due to neutralized stellar wind ions and evaporating helium atoms from the planet can account, at least qualitatively, for the broad line profile observed during \nuno\, corresponding to a velocity in excess of $\sim 40$~km~s$^{-1}$, and its remarkable blueshift (cf. Fig.~\ref{additional}, top panel). On the other hand, the reduced blueshifts and/or smaller line widths during the other nights (cf. Fig.~\ref{additional}, bottom panels) can be accounted for by a significantly slower stellar wind. It is interesting to note that not only the average stellar wind velocity is relevant here to account for the blueshift of the line centroids, but also its velocity dispersion close to the planet, owing to the turbulence in the wind itself, that plays a role in the line broadening because the wind neutrals collide with the helium atoms of the planetary evaporating flow and exchange momentum with them.
                
                \section{Plasma stratification inside the planetary magnetosphere}
                \label{plasma_distr}
                
                The plasma evaporating from the planet and filling its magnetosphere reaches a condition of mechanical equilibrium described by the equation
                \begin{equation}
                        -\nabla p + \rho \nabla \Phi + {\vec J} \times {\vec B}= 0,
                        \label{mech_eq}
                \end{equation}
                where $p$ is the plasma pressure, $\rho$ its density, $\Phi$ the gravitational potential, $\vec J$ the current density, and $\vec B$ the magnetic field. The centrifugal force can be neglected because the planet is rotating slowly (cf. Appendix~\ref{appendix_wind}). Therefore, the gravitational potential can be written as
                \begin{equation}
                        \Phi = \frac{GM_{\rm p}}{r},
                \end{equation}
                where $G$ is the gravitation constant, $M_{\rm p}$ the mass of the planet, and $r$ the radial distance from the centre of the planet. We can project Eq.~\eqref{mech_eq} along any magnetic field line having a unit vector tangent $\hat{\vec s}$, that is, ${\vec B} = B \,\hat{\vec s}$, yielding
                \begin{equation}
                        - \frac{1}{\rho}\frac{\partial p}{\partial s} - \frac{GM_{\rm p}}{r^{2}} \hat{\vec r} \cdot \hat{\vec s} = 0.  
                        \label{mech_eq1}
                \end{equation}
                The scalar product of the unit vectors $\hat{\vec r} \cdot \hat{\vec s} = \cos \psi$, where $\psi$ is the angle between the tangent to the magnetic field line and the radial direction. Hence $dr = ds \cos \psi$ and Eq.~\eqref{mech_eq1} becomes
                \begin{equation}
                        -\frac{1}{\rho}\frac{\partial p}{\partial r} -\frac{GM_{\rm p}}{r^{2}} = 0.
                        \label{mech_eq2}
                \end{equation}
                Adopting an isothermal atmosphere with a uniform temperature $T$, the ideal gas law gives $p = \tilde{R}T \rho/\tilde{\mu}$, where $\tilde{R}$ is the gas constant and $\tilde{\mu}$ the mean molecular weight of the plasma. For a fully ionized, solar composition plasma, $\tilde{\mu} \sim 0.6$.  Equation~\eqref{mech_eq2} can be integrated analytically making use of the ideal gas law and introducing the pressure scale height $H_{\rm p} \equiv \tilde{R}T/(\tilde{\mu} g)$, where $g = GM_{\rm p}/R_{\rm p}^{2}$ is the acceleration of gravity at the surface of the planet
                \begin{equation}
                        \rho(r) = \rho_{0} \exp\left[ \left( \frac{R_{\rm p}}{H_{\rm p}} \right) \left( \frac{R_{\rm p}}{r} -1 \right) \right], \mbox{~$r \geq R_{\rm p}$}, 
                        \label{dens_run}
                \end{equation}
                where $\rho_{0}$ is the density at the surface of the planet, where $r  = R_{\rm p}$, and that coincides with the base of the magnetosphere in our simple model. 
                
                A lower limit for the density in the magnetosphere can be found from Eq.~\eqref{dens_run} in the limit $r \rightarrow \infty$, obtaining $\rho > \rho_{0} \exp(-R_{\rm p}/H_{\rm p})$. In the case of V1298~Tau~b, assuming a solar composition plasma and an isothermal magnetosphere with $T= 10^{4}$~K, we obtain $H_{\rm p} \sim 1.6 \, R_{\rm p}$ giving a lower limit for the density inside the magnetosphere of $0.53 \rho_{0}$. In other words, the density does not vary by more than a factor of $\sim 2$ inside the planetary magnetosphere in our isothermal model. This is essentially a consequence of the high planet radial inflation yielding a very small surface gravity and therefore a large pressure scale height. On the other hand, for a planet having a stronger surface gravity $g$ and the same temperature of the magnetospheric plasma, the pressure scale height $H_{\rm p} \propto g^{-1}$ will be smaller and the reduction of the density at large distance from the planet surface will be remarkably larger because the limit density decreases as $\rho_{0}\exp (-R_{\rm p}/H_{\rm p})$.

        \end{appendix}
        
\end{document}